\documentclass[twocolumn,english,prb,aps,showpacs,superscriptaddress]{revtex4-1}
\usepackage{amsmath}
\usepackage{amssymb}
\usepackage{graphicx}
\usepackage{esint}
\makeatletter

 \@ifundefined{textcolor}{}
 {%
   \definecolor{BLACK}{gray}{0}
   \definecolor{WHITE}{gray}{1}
   \definecolor{RED}{rgb}{1,0,0}
   \definecolor{GREEN}{rgb}{0,1,0}
   \definecolor{BLUE}{rgb}{0,0,1}
   \definecolor{CYAN}{cmyk}{1,0,0,0}
   \definecolor{MAGENTA}{cmyk}{0,1,0,0}
   \definecolor{YELLOW}{cmyk}{0,0,1,0}
 }

\usepackage{dcolumn}% Align table columns on decimal point
\usepackage{bm}% bold math
\makeatother

\begin{document}

\title{Perturbation theory for an Anderson quantum dot asymmetrically attached
to two superconducting leads}

\author{M. \v{Z}onda}

\affiliation{Department of Condensed Matter Physics, Faculty of Mathematics and
Physics, Charles University in Prague, Ke Karlovu 5, CZ-12116 Praha
2, Czech Republic}

\author{V. Pokorn\'{y}}

\affiliation{Institute of Physics, Academy of Sciences of the Czech Republic,
Na Slovance 2, CZ-18221 Praha 8, Czech Republic}

\affiliation{Theoretical Physics III, Center for Electronic Correlations and Magnetism,
Institute of Physics, University of Augsburg, D-86135 Augsburg, Germany}

\author{V. Jani\v{s}}

\affiliation{Institute of Physics, Academy of Sciences of the Czech Republic,
Na Slovance 2, CZ-18221 Praha 8, Czech Republic}

%\affiliation{Fulbright Scholar at Department of Physics and Astronomy, Louisiana
%State University, Baton Rouge, LA 70803, USA}

\author{T. Novotn\'{y}}

\affiliation{Department of Condensed Matter Physics, Faculty of Mathematics and
Physics, Charles University in Prague, Ke Karlovu 5, CZ-12116 Praha
2, Czech Republic}
\email{tno@karlov.mff.cuni.cz}

\date{\today}
\begin{abstract}
Self-consistent perturbation expansion up to the second order in the
interaction strength is used to study a single-level quantum dot with
local Coulomb repulsion attached asymmetrically to two generally different
superconducting leads. At zero temperature and wide range of other
parameters the spin-symmetric version of the expansion yields excellent
results for the position of the $0-\pi$ impurity quantum phase transition
boundary and Josephson current together with the energy of Andreev
bound states in the $0$-phase as confirmed by numerical
calculations using the Numerical Renormalisation Group method. We
analytically prove that the method is charge-conserving as well as
thermodynamically consistent. Explicit formulas for the position of
the $0-\pi$ phase boundary are presented for the Hartree-Fock approximation
as well as for its variant called Generalized Atomic Limit. It
is shown that the Generalized Atomic Limit can be used as a quick estimate
for the position of the phase boundary at half-filling in a broad
range of parameters. We apply our second order perturbation method
to the interpretation of the existing experimental data on the phase
boundary with very satisfactory outcome suggesting that the so far employed
heavy numerical tools such as Numerical Renormalization Group and/or
Quantum Monte Carlo are not necessary in a class of generic situations and
can be safely replaced by a perturbative approach. 
\end{abstract}


\maketitle

\section{Introduction}

In the last decade advances in the fabrication of nano-devices enabled
to connect quantum dots with superconducting (SC) leads forming superconducting
quantum dot nanostructures generalizing the conventional Josephson
junctions \cite{DeFranceschi10}. Many experimental realizations of
this concept using various BCS materials for the superconducting leads
(\emph{Al}, \emph{Pb}, or \emph{Nb}) and a great variety of quantum
dots formed in semiconducting nanowires\cite{vanDam06,Jesper13} or
dots,\cite{Katsaros10} carbon nanotubes,\cite{Kasumov99,Morpurgo99,Kasumov03,Jarillo06,Jorgensen06,Cleuziou06,Jorgensen07,Grove07,Pallecchi08,Zhang08,Jorgensen09,Eichler09,Liu09,Pillet10,Lee12,Maurand12,Pillet13,Kumar14,Delagrange15}
or even single $C_{60}$ molecules\cite{Winkelmann09} demonstrate
the versatility of such setups. Major advantage of the superconducting
quantum dots over conventional microscopic Josephson junctions lies
in the possibility of a detailed control of their microscopic parameters,
e.g., by tuning the onsite energy by a gate voltage. Such a high tunability
is promising for potential applications of these hybrids in the nanoelectronics
(e.g., as a superconducting single-electron transistor) or quantum
computing as well as for detailed studies of their non-trivial physical
properties. 

These include Josephson supercurrent, Andreev subgap transport
and the way they are influenced by the zero-dimensional nature of
the superconducting quantum dots with finite-size quantized levels
and potentially strong effects of the local Coulomb interaction leading
to strongly correlated phenomena such as the Kondo effect.\cite{Yeyati11}
In many cases the system can be very well described by a simplest
single impurity Anderson model (SIAM) coupled to BCS leads,\cite{Luitz12}
which, depending on particular parameters, may exhibit so called $0-\pi$
transition signaled by the sign reversal of the supercurrent.\cite{vanDam06,Cleuziou06,Jorgensen07,Eichler09,Maurand12,Delagrange15}
The $0-\pi$ transition is induced by the underlying impurity quantum
phase transition (QPT) related to the crossing of the lowest many-body
eigenstates of the system from a spin-singlet ground state with positive
supercurrent ($0$-phase) to a spin-doublet state with negative supercurrent
($\pi$-phase).\cite{Matsuura77,Glazman89,Rozhkov99,Yoshioka00,Siano04,Choi04,Sellier05,Novotny05,Karrasch08,Meng09,Luitz12}
In single-particle spectral properties this transition is associated
with crossing of the Andreev bound states (ABS) at the Fermi energy
as has also been observed experimentally.\cite{Pillet10,Pillet13,Jesper13}

A number of theoretical techniques have been used to address the $0-\pi$
transition and related properties of superconducting quantum dots.
A very good quantitative agreement with the experiments\cite{Luitz12,Pillet13,Delagrange15}
can be obtained in a wide range of parameters using heavy numerics
such as numerical renormalization group (NRG)\cite{Yoshioka00,Choi04,Bauer07,Tanaka07,Hecht08,Oguri04,Rodero12,Oguri13,Pillet13}
and quantum Monte Carlo (QMC).\cite{Siano04,Luitz10,Luitz12,Delagrange15}
However, both NRG and QMC are time and computational resources consuming.
Alternative (semi)analytical methods based on various, often quite
sophisticated, perturbation approaches either around non-interacting
limit ($U=0$) such as the mean-field theory,\cite{Rozhkov99,Yoshioka00,Vecino03,Rodero12}
slave particles,\cite{Clerk00,Sellier05} and functional renormalization
group (fRG)\cite{Karrasch08,Karrasch-PhD10} or around the atomic
limit ($\Gamma\to0$ or $\Delta\to\infty$)\cite{Konig08,Meng09,Droste12}
have been used for qualitative and in some limits even a quantitative
description of the superconducting quantum dot properties. Yet, none
of the mentioned methods with the exception of the mean-field/Hatree-Fock
(HF) approximation are sufficiently simple and at the same time versatile
to serve as a generic (semi)analytical solver. HF approximation has
the attributes of the generic method,\cite{Rodero12} yet, it suffers
from fundamental conceptual problems, namely, it identifies the
$0-\pi$ transition with the point of breaking of the spin-symmetric
solution and attributes the $\pi$ phase to the magnetic solution
of the self-consistent HF equations. This unphysical breaking of the
spin symmetry together with the ensuing discontinuities of various physical
quantities even at non-zero temperatures contradicting the experimental
observations disqualify the unrestricted HF approach as a reliable
solver for the superconducting SIAM. 

\begin{figure}
\includegraphics[width=0.9\columnwidth]{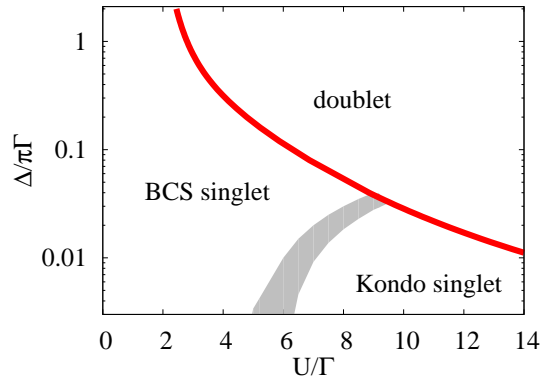}\caption{(Color online) Sketch of the phase diagram in the $U-\Delta$ plane
of the superconducting single-impurity Anderson model with symmetric
leads at half-filling. Full red line separates the singlet and doublet
ground states and the shade region signals the crossover between the
two kinds of singlet ground states. The perturbative approach presented
in this paper works well in the whole BCS singlet regime as demonstrated
in detail in Figs.~\ref{fig:PhD_U_delta} and \ref{fig:One-Particle-Quantities}
below. \label{fig:Phase-diagram-Intro}}
\end{figure}

Surprisingly, with the exception of a few fragmented precursors,\cite{Vecino03,Meng-master09}
it has not been noticed until very recently\cite{Zonda15,Janis15}
that the resummed perturbation theory incorporating second-order
dynamical corrections to the spin-symmetric HF solution yields at
zero temperature a nearly perfect description of the $0$ phase for
symmetric leads in a wide range of parameters. The aim of this
work is to demonstrate that second-order perturbation
theory is an efficient and reliable method not only for the symmetric
leads, but also for a more general and  realistic case of asymmetric
tunnel coupling to different  leads  (i.e., with
various values of superconducting gaps). This method is numerically
much less expensive than NRG or QMC. 
Note that in the general case one has to deal with the two-channel Anderson model, 
therefore, the introduced second-order perturbation theory can be 10 or even 100 times faster 
(depending on parameters and used CPU cores) than the fully convergent NRG calculations. 
Simultaneously, this method gives nearly
perfect results for the physical quantities in the $0$-phase at zero
temperature in a wide range of parameters corresponding to weakly-
and intermediately correlated regime, where the conventional deployment
of NRG is unnecessary. As known from previous studies\cite{Yoshioka00,Bauer07}
the ground-state in this regime is the BCS singlet in contrast to
the strongly correlated regime where the ground state is the Kondo singlet.
We illustrate this in Fig.~\ref{fig:Phase-diagram-Intro} which depicts
the ground state phase diagram in the $U-\Delta$ plane for quantum
dots with symmetric leads at half-filling. The crossover region between
the BCS and Kondo singlets is approximately plotted as a gray stripe.
The BCS singlet regime covers a broad range of parameters (note the
logarithmic scale on the vertical axis) where second-order perturbation
theory is in a nearly perfect agreement with NRG. Thus, we advocate
this method to be the generic first-choice solver for the properties
of the $0$ phase. 

In order to support this standpoint we carefully examined
formal properties of the approximation such as charge conservation,
gauge invariance, and thermodynamic consistency and showed that it
preserves all these important requirements.\footnote{Its simplified version (DC) used for most of our numerical calculations strictly satisfies the charge conservation only for equal SC gaps and marginally breaks it for the general case, for details see~Sec.~\ref{sub:Applicability} and Appendix~\ref{App:Charge-conservation}}
Then, we systematically studied the zero-temperature $0$-phase
quantities in a wide parameter range, paying a special attention to
the position of the phase boundary between the $0$ and $\pi$ phases.
We identified the limits of applicability of our method by direct
comparison with NRG data obtained via the NRG Ljubljana open source
code.\cite{NRGLjubljana,Zitko09} At small enough temperatures, the
boundary depends only weakly on temperature\cite{Karrasch08,Luitz10}
and, therefore, our zero-temperature results are directly applicable
to the existing experimental data. We finally compared our predictions
with two experiments with excellent agreement, further justifying our
claims. 

The outline of the paper is as follows. In the next Sec.~\ref{sec:Theory}
we introduce  the superconducting SIAM,
the Matsubara Green function methodology of the perturbation theory
together with the Josephson current, and ABS formulas in the first subsection, while in the
second part we study charge conservation, gauge invariance, and
thermodynamic consistency conditions to be obeyed by approximations.
In Sec.~\ref{sec:Approximations}, we introduce and analyze properties of the Hartree-Fock approximation 
and its dynamical corrections. In the following Sec.~\ref{sec:Results},
after a brief summary of technical issues concerning the evaluation
of the approximative equations, we present results for the position
of the $0-\pi$ boundary first for the case of identical leads (equal
SC gaps) and then for different leads with unequal gaps. Finally,
in the last subsection we discuss in details the applicability and
limitations of our method as demonstrated on various single-particle
quantities in the $0$ phase, such as ABS energies and/or induced SC
gap. In Sec.~\ref{sec:Experiments}, we present  comparison of quantitative results of our theory with two existing experiments
on the position of the $0-\pi$ boundary. We conclude our work in
the last Sec.~\ref{sec:Conclusions}. Supporting technical calculations
for the HF boundary and charge conservation in the second-order perturbation
theory are deferred to Appendixes \ref{App:Hartree-Fock} and \ref{App:Charge-conservation}.

\section{Theory and methods\label{sec:Theory}}

\subsection{Model and observables}
\begin{subequations}
We use a single-impurity Anderson model\cite{Choi04,Siano04,Luitz12,Pillet13}
as a model of the quantum dot with well-separated energy levels connected
to two asymmetric superconducting leads.\cite{Tanaka07} The Hamiltonian
of the system is given by 
\begin{equation}
\mathcal{H}=\mathcal{H}_{{\rm dot}}+\sum_{\alpha=R,L}(\mathcal{H}_{{\rm lead}}^{\alpha}+\mathcal{H}_{T}^{\alpha}).\label{eq:Ham}
\end{equation}
The first term represents a single impurity with the level energy
$\varepsilon$ and the local Coulomb repulsion $U$: 
\begin{equation}
\mathcal{H}_{{\rm dot}}=\varepsilon\sum_{\sigma=\uparrow,\downarrow}d_{\sigma}^{\dagger}d_{\sigma}+Ud_{\uparrow}^{\dagger}d_{\uparrow}d_{\downarrow}^{\dagger}d_{\downarrow},
\end{equation}
where $d_{\sigma}^{\dagger}$ ($d_{\sigma}$) creates (annihilates)
an electron with the spin $\sigma$ on the impurity. The second term
of Eq.~\eqref{eq:Ham} is the BCS Hamiltonian of the leads

\begin{equation}
\mathcal{H}_{{\rm lead}}^{\alpha}=\sum_{\mathbf{k}\sigma}\epsilon_{\alpha}(\mathbf{k})c_{\alpha\mathbf{k}\sigma}^{\dagger}c_{\alpha\mathbf{k}\sigma}-\Delta_{\alpha}\sum_{\mathbf{k}}(e^{i\Phi_{\alpha}}c_{\alpha\mathbf{k}\uparrow}^{\dagger}c_{\alpha\mathbf{\ -k}\downarrow}^{\dagger}+\textrm{H.c.}),
\end{equation}
with $\alpha=L,R$ denoting the left and right leads and $c_{\alpha\mathbf{k}\sigma}^{\dagger}$
($c_{\alpha\mathbf{k}\sigma}$) creating (annihilating) conduction
electrons. Finally, the hybridization term between the impurity and
the contacts is given by 
\begin{equation}
\mathcal{H}_{T}^{\alpha}=-t_{\alpha}\sum_{\mathbf{k}\sigma}(c_{\alpha\mathbf{k}\sigma}^{\dagger}d_{\sigma}+\textrm{H.c.}).
\end{equation}

\end{subequations}

All the studied quantities can be expressed with the help of the impurity
one-electron (imaginary time/Matsubara) Green function (GF), which
is a $2\times2$ matrix in the Nambu spinor formalism 
\begin{equation}
\begin{aligned}\widehat{G}_{\sigma}(\tau-\tau') & \equiv\begin{pmatrix}G_{\sigma}(\tau-\tau')\ , & \mathcal{G}_{-\sigma}(\tau-\tau')\\
\bar{\mathcal{G}}_{\sigma}(\tau-\tau')\ , & \bar{G}_{-\sigma}(\tau-\tau')
\end{pmatrix}\\
= & -\begin{pmatrix}\langle\mathbb{T}[d_{\sigma}(\tau)d_{\sigma}^{\dagger}(\tau')]\rangle\ , & \langle\mathbb{T}[d_{\sigma}(\tau)d_{-\sigma}(\tau')]\rangle\\[0.3em]
\langle\mathbb{T}[d_{-\sigma}^{\dagger}(\tau)d_{\sigma}^{\dagger}(\tau')]\rangle\ , & \langle\mathbb{T}[d_{-\sigma}^{\dagger}(\tau)d_{-\sigma}(\tau')]\rangle
\end{pmatrix}\\
= & \begin{pmatrix}\quad\includegraphics[width=0.15\textwidth]{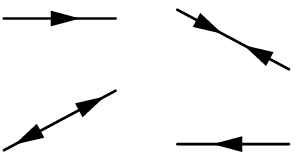}\quad\end{pmatrix},
\end{aligned}
\label{eq:GFdef}
\end{equation}
where the bar denotes the hole function. Since we only consider spin-symmetric
solutions throughout the whole paper, we skip the spin index and we
also set $e=\hbar=1$ from now on. The exact form of the unperturbed
($U=0$) impurity GF can be written as a function of Matsubara frequencies
$\omega_{n}\equiv(2n+1)\pi/\beta$ (see Appendix A of Ref.~\onlinecite{Novotny05}): 
\begin{subequations}
\begin{equation}
\widehat{G}_{0}(i\omega_{n})=\begin{pmatrix}i\omega_{n}[1+s(i\omega_{n})]-\varepsilon\ , & \Delta_{\Phi}(i\omega_{n})\\[0.3em]
\Delta_{\Phi}^{*}(i\omega_{n})\ , & i\omega_{n}[1+s(i\omega_{n})]+\varepsilon
\end{pmatrix}^{-1},\label{eq:D0}
\end{equation}
where 
\begin{equation}
s(i\omega_{n})=\sum_{\alpha=L,R}\frac{\Gamma_{\alpha}}{\sqrt{\Delta_{\alpha}^{2}+\omega_{n}^{2}}}\label{eq:D0abb1}
\end{equation}
is the hybridization self-energy due to the coupling of the impurity
to the superconducting leads. We have denoted by $\Gamma_{L,R}=\pi t_{L,R}^{2}\rho_{L,R}$
the normal-state tunnel-coupling magnitude (with $\rho_{L,R}$ being
the normal-state density of states of the respective lead electrons
at the Fermi energy) and 
\begin{equation}
\Delta_{\Phi}(i\omega_{n})=\sum_{\alpha=L,R}\frac{\Gamma_{\alpha}\Delta_{\alpha}}{\sqrt{\Delta_{\alpha}^{2}+\omega_{n}^{2}}}e^{i\Phi_{\alpha}}.\label{eq:D0abb2}
\end{equation}

\end{subequations}

The impact of the Coulomb repulsion $U>0$ on the Green function is
included in the interaction self-energy matrix $\widehat{\Sigma}(i\omega_{n})\equiv\left(\begin{smallmatrix}\Sigma(i\omega_{n}), & \mathcal{S}(i\omega_{n})\\
\mathcal{\bar{S}}(i\omega_{n}), & \bar{\Sigma}(i\omega_{n})
\end{smallmatrix}\right)$, so that the full propagator in the spin-symmetric situation is determined
by the Dyson equation $\widehat{G}^{-1}(i\omega_{n})=\widehat{G}_{0}^{-1}(i\omega_{n})-\widehat{\Sigma}(i\omega_{n})$.
Symmetry relations for the spin-symmetric version of the Green
function, Eq.~\eqref{eq:GFdef}, reformulated in the Matsubara
representation\cite[Sec.~9.3.3]{Karrasch-PhD10} are $\bar{G}(i\omega_{n})=-G^{*}(i\omega_{n})=-G(-i\omega_{n})$
and $\mathcal{\bar{G}}(i\omega_{n})=\mathcal{G}^{*}(i\omega_{n})=\mathcal{G}^{*}(-i\omega_{n})$.
The same applies to the self-energies, i.e., $\bar{\Sigma}(i\omega_{n})=-\Sigma^{*}(i\omega_{n})=-\Sigma(-i\omega_{n}),\,\bar{\mathcal{S}}(i\omega_{n})~=\mathcal{S}^{*}(i\omega_{n})=\mathcal{S}^{*}(-i\omega_{n})$.
Here the asterisk stands for time inversion being complex conjugation
in the Matsubara formalism. Consequently, the interacting Green function
explicitly reads as 
\begin{widetext}
\begin{gather}
\widehat{G}(i\omega_{n})=-\frac{1}{D(i\omega_{n})}\begin{pmatrix}i\omega_{n}[1+s(i\omega_{n})]+\varepsilon+\Sigma^{*}(i\omega_{n}), & -\Delta_{\Phi}(i\omega_{n})+\mathcal{S}(i\omega_{n})\\[0.3em]
-\Delta_{\Phi}^{*}(i\omega_{n})+\mathcal{S}^{*}(i\omega_{n}), & i\omega_{n}[1+s(i\omega_{n})]-\varepsilon-\Sigma(i\omega_{n})
\end{pmatrix}.\label{eq:GF}
\end{gather}
The existence and the position of the ABS are determined by zeros
of the negative determinant of the inverse Green function (i.e., poles
of the Green function) 

\begin{gather}
\begin{split}D(i\omega_{n}) & \equiv-\det[\widehat{G}^{-1}(i\omega_{n})]=\omega_{n}^{2}\left[1+s(i\omega_{n})\right]^{2}+\left|\varepsilon+\Sigma(i\omega_{n})\right|^{2}+\left|\Delta_{\Phi}(i\omega_{n})-\mathcal{S}(i\omega_{n})\right|^{2}\\
 & =D(-i\omega_{n})=D^{*}(i\omega_{n})\geq0\,.
\end{split}
\label{eq:Det}
\end{gather}

\end{widetext}

The determinant analytically continued from Matsubara to the real
frequency axis is real within the gap and can go through zero $D(\omega_{0})=0$
determining the (real) in-gap energies $\pm\omega_{0}$ of the ABS
symmetrically placed around the Fermi energy lying in the middle of
the gap. ABS are crucial for transport of the Cooper pairs through
the quantum dot because they usually provide the dominant contribution
to the dissipation-less Josephson current through the impurity. Furthermore,
their crossing at the Fermi energy determines the phase boundary between
$0$- and $\pi$ phases as their zero energy is equivalent to the
degeneracy of the two lowest-energy many-body eigenstates of the system,
which is the point of the impurity QPT.\cite{Pillet13}

The Josephson current out of the dot into the respective reservoir
$J_{\alpha}$ is defined from the Heisenberg equation of motion for
the particle number in the reservoir $J_{\alpha}\equiv d\left\langle \sum_{\mathbf{k}}c_{\alpha\mathbf{k}\sigma}^{\dagger}c_{\alpha\mathbf{k}\sigma}^{\phantom{\dagger}}\right\rangle /dt=-i\left\langle \left[\sum_{\mathbf{k}}c_{\alpha\mathbf{k}\sigma}^{\dagger}c_{\alpha\mathbf{k}\sigma}^{\phantom{\dagger}},\mathcal{H}\right]\right\rangle $
and can be evaluated as a Matsubara sum of the anomalous Green function
\begin{equation}
\begin{split}J_{\alpha} & =\frac{2}{\beta}\Im\sum_{\omega_{n}}\frac{\Gamma_{\alpha}}{\sqrt{\Delta_{\alpha}^{2}+\omega_{n}^{2}}}\\
 & \times\mathrm{Tr}\left[\begin{pmatrix}0 & -\Delta_{\alpha}e^{i\Phi_{\alpha}}\\
\Delta_{\alpha}e^{-i\Phi_{\alpha}} & 0
\end{pmatrix}\widehat{G}(i\omega_{n})\right]\\
 & =\frac{2}{\beta}\Im\sum_{\omega_{n}}\frac{\Gamma_{\alpha}\Delta_{\alpha}}{\sqrt{\Delta_{\alpha}^{2}+\omega_{n}^{2}}}\left[\mathcal{G}(i\omega_{n})e^{-i\Phi_{\alpha}}-\mathcal{\bar{G}}(i\omega_{n})e^{i\Phi_{\alpha}}\right]\\
 & =\frac{4}{\beta}\sum_{\omega_{n}}\frac{\Gamma_{\alpha}\Delta_{\alpha}}{\sqrt{\Delta_{\alpha}^{2}+\omega_{n}^{2}}}\Im\left[\mathcal{G}(i\omega_{n})e^{-i\Phi_{\alpha}}\right],
\end{split}
\label{eq:JC}
\end{equation}
where $\alpha=L,R$ as before. For any approximative treatment such
as our perturbation expansion in $U$ there always arises an important
question of charge conservation, i.e., whether $J_{L}=-J_{R}$ and
thermodynamic consistency, i.e., whether Josephson current calculated
by different approaches, e.g., directly from Eq.~\eqref{eq:JC} or
as a phase-derivative of the associated free energy, gives the same.
We devote  the following subsection to these nontrivial fundamental questions.

\subsection{Charge conservation, thermodynamic consistency, and gauge invariance\label{sub:charge-conservation}}

Any consistent approximation must respect charge conservation, i.e.,
the Josephson currents through the left and right interfaces sum up
to zero (due to the above convention for the definition of the Josephson
current as flowing \emph{into} the respective lead). The condition
$J_{L}+J_{R}=0$ can be rewritten with the help of the second line
of Eq.~\eqref{eq:JC} as 
\begin{equation}
\begin{split}0 & =\frac{2}{\beta}\Im\sum_{\omega_{n},\alpha}\frac{\Gamma_{\alpha}\Delta_{\alpha}}{\sqrt{\Delta_{\alpha}^{2}+\omega_{n}^{2}}}\left[\mathcal{G}(i\omega_{n})e^{-i\Phi_{\alpha}}-\mathcal{G}^{*}(i\omega_{n})e^{i\Phi_{\alpha}}\right]\\
 & =\frac{2}{\beta}\Im\sum_{\omega_{n}}\left[\Delta_{\Phi}^{*}(i\omega_{n})\mathcal{G}(i\omega_{n})-\Delta_{\Phi}(i\omega_{n})\mathcal{G}^{*}(i\omega_{n})\right]\\
 & =\frac{2}{\beta}\Im\sum_{\omega_{n}}\left[\mathcal{S}^{*}(i\omega_{n})\mathcal{G}(i\omega_{n})-\mathcal{S}(i\omega_{n})\mathcal{G}^{*}(i\omega_{n})\right]\\
 & =\frac{4}{\beta}\Im\sum_{\omega_{n}}\mathcal{S}^{*}(i\omega_{n})\mathcal{G}(i\omega_{n})%=2\Im\sum_{\omega_{n}}\mathcal{S}^{*}(-i\omega_{n})\mathcal{G}(-i\omega_{n})
\,,
\end{split}
\label{eq:charge-cons}
\end{equation}
where we used $\Delta_{\Phi}(i\omega_{n})=D(i\omega_{n})\mathcal{G}(i\omega_{n})+\mathcal{S}(i\omega_{n})$ from Eq.~\eqref{eq:GF} and reality of $D(i\omega_{n})$ from Eq.~\eqref{eq:Det}. For
symmetric leads one can choose real $\Delta_{\Phi}(i\omega_{n})$
and consequently the self-energy and Green function fulfill $\mathcal{S}^{*}(i\omega_{n})=\mathcal{S}(-i\omega_{n})$
and $\mathcal{G}^{*}(i\omega_{n})=\mathcal{G}(-i\omega_{n})$. The
charge conservation condition \eqref{eq:charge-cons} is then satisfied
automatically as can be seen by the change of the summation variable
$\omega_{n}\to-\omega_{n}$. Since for asymmetric leads one cannot
guarantee the reality of $\Delta_{\Phi}(i\omega_{n})$ for all frequencies
and thus $\mathcal{S}^{*}(i\omega_{n})\neq\mathcal{S}(-i\omega_{n})$
approximations must be checked for fulfilling charge conservation
\eqref{eq:charge-cons} by explicit verification. 

Apart from a direct approach to charge conservation via the explicit
formula for the Josephson current, Eq.~\eqref{eq:JC}, we may also employ
an indirect one starting with the phase-dependent grand potential
(``free energy'') of the system. The dissipationless Josephson
current can also be determined as the phase derivative of the thermodynamic
potential. Approximate calculations of a thermodynamic quantity (such
as the Josephson current here) lead to the same result when different
equivalent representations are used only in \emph{thermodynamically
consistent} approaches in the Baym sense.\cite{Baym61,Baym62}

A thermodynamically consistent approximation can be generated from
a Luttinger-Ward functional $\phi[\widehat{G}]$. It is represented
in terms of the full one-electron Green function $\widehat{G}$, Eq.~\eqref{eq:GF},
from which the self-energy is determined via a functional derivative.
In our case with asymmetric leads we have to treat both electron and
hole variables as independent parameters. Hence, the functional derivates
determining the self-energies read as $\Sigma(i\omega_{n})=\beta\delta\phi[\widehat{G}]/\delta G^{*}(-i\omega_{n})$
for the normal part and $\mathcal{S}(i\omega_{n})=\beta\delta\phi[\widehat{G}]/\delta\mathcal{G}^{*}(-i\omega_{n})$
for the anomalous one and analogously for the self-energies $\Sigma^{*}(i\omega_{n})$
and $\mathcal{S}^{*}(i\omega_{n})$. The grand potential then contains
both electron and hole variables, where the hole variables are decorated
with asterisks that have the meaning of complex conjugation (only)
in equilibrium. The grand potential can be represented with the aid of the Luttinger-Ward
functional as follows: 
\begin{widetext}
\begin{multline}
2\Omega[\widehat{G},\widehat{\Sigma}]= \phi[\widehat{G}]-\frac{1}{\beta}\sum_{\omega_{n}}e^{i\omega_{n}0^{+}}\left\{ G(i\omega_{n})\Sigma^{*}(-i\omega_{n})+\ G^{*}(-i\omega_{n})\Sigma(i\omega_{n})+\mathcal{G}(i\omega_{n})\mathcal{S}^{*}(-i\omega_{n})+\mathcal{G}^{*}(-i\omega_{n})\mathcal{S}(i\omega_{n})\phantom{\frac{1}{2}}\right.\\
\left.+\log\left[[i\omega_{n}(1+s(i\omega_{n}))-\varepsilon-\Sigma(i\omega_{n})][i\omega_{n}(1+s(i\omega_{n}))+\varepsilon+\Sigma^{*}(i\omega_{n})]-[\Delta_{\Phi}(i\omega_{n})-\mathcal{S}(i\omega_{n})][\Delta_{\Phi}^{*}(i\omega_{n})-\mathcal{S}^{*}(i\omega_{n})]\right]\right\}\,.
\end{multline}

\end{widetext}

Complex variables $G(i\omega_{n}),G^{*}(i\omega_{n})$, $\Sigma(i\omega_{n}),\Sigma^{*}(i\omega_{n})$
as well as $\mathcal{G}(i\omega_{n}),\mathcal{G}^{*}(i\omega_{n})$,
$\mathcal{S}(i\omega_{n}),\mathcal{S}^{*}(i\omega_{n})$ are variational
parameters the physical values of which are determined from the stationarity
of the grand potential $\Omega[\widehat{G},\widehat{\Sigma}]$. Due
to the electron-hole symmetry the electron and hole contributions
to the thermodynamic potential are identical, hence, the factor $2$
on the left-hand side. An approximation is thermodynamically consistent
and conserving if we are able to determine explicitly the Luttinger-Ward
functional $\phi[\widehat{G}]$. Such approximations are called $\phi$ derivable.

Reliable and physically acceptable approximations should not only
be thermodynamically consistent but must be gauge invariant. Observables
in Josephson junction setups depend on the phase difference between
the leads but they cannot depend on the absolute values of the two
phases. In other words, the physics must be invariant with respect
to a global phase shift $\Phi_{L,R}\mapsto\Phi_{L,R}+\Delta\Phi$,
which is a manifestation of gauge invariance. Obviously, the building
elements of the theory, the Green functions, Eqs.~\eqref{eq:D0}
and~\eqref{eq:GF}, are not invariant and we must always check that
they enter the measurable quantities, such as the supercurrent \eqref{eq:JC},
in a way that preserves gauge invariance. The resulting self-energy
$\widehat{\Sigma}(i\omega_{n})$ and consequently $\widehat{G}(i\omega_{n})$
transform equally as $\widehat{G}_{0}(i\omega_{n})$ \eqref{eq:D0}
under the gauge transformation in thermodynamically consistent approximations.
That is, only the off-diagonal elements pick up the global phase shift
(with the respective sign) from which we can immediately see that
the Josephson current \eqref{eq:JC} is indeed gauge-invariant as
it should be.

We can also use the functional of the grand potential $\Omega[\widehat{G},\widehat{\Sigma}]$
to prove gauge invariance of a $\phi$-derivable approximation. In
gauge-invariant theories, thermodynamic potentials depend only on the
phase difference, i.e., $\Omega[\widehat{G},\widehat{\Sigma}](\Phi_{L},\Phi_{R})\equiv\Omega[\widehat{G},\widehat{\Sigma}](\Phi_{L}-\Phi_{R})$.
Consequently, charge conservation $J_{L}\equiv\frac{\partial\Omega\left[\widehat{G},\widehat{\Sigma}\right](\Phi_{L}-\Phi_{R})}{\partial\Phi_{L}}=-\frac{\partial\Omega\left[\widehat{G},\widehat{\Sigma}\right](\Phi_{L}-\Phi_{R})}{\partial\Phi_{R}}\equiv-J_{R}$
immediately follows.

\section{Approximation schemes\label{sec:Approximations}}

Since the exact expression for the self-energy of the superconducting
Anderson impurity model is unknown, we have applied the standard Matsubara
resummed perturbation theory in the interaction strength $U$ to summing
up one-particle-irreducible diagrams for the self-energy in terms
of the dressed one-particle Green function. To avoid the unphysical
spin polarization of the impurity, which can be easily obtained within
the resummed (dressed) approach from the self-consistent solution
of a nonlinear equation for the single-particle Green function, we
restrict the solution to the spin-symmetric case. This results in
the situation when for certain parameters at zero temperature there
exists no solution for the Green function. The breakdown of the solution
coincides with the crossing of ABS energies at the Fermi energy, i.e.,
with the $0-\pi$ quantum phase transition. Thus, while the $0$-phase,
which can be smoothly connected with the noninteracting case $U=0$,
can be captured by the perturbative approach, the $\pi$ phase with
doubly degenerate ground state is beyond the reach of this simple perturbation
theory. We explicitly demonstrate this concept for the Hartree-Fock
solution, where all quantities at QPT can be addressed analytically,
but the same scheme carries over to higher orders of the perturbation
theory, in particular to second order, being  the main focus
of our study. The very possibility of a description
of the $\pi$-phase within a (suitably modified) perturbative approach
remains an open question as we discuss in more detail later in Sec.~\ref{sub:Applicability}.

\subsection{Spin-symmetric (restricted) Hartree-Fock approximation \label{HFA}}
\begin{subequations}
Mathematical expressions for first-order perturbation expansion in
$U$, Hartree-Fock (HF) contributions, read 
\begin{align}
\Sigma^{\mathrm{HF}} & =\frac{U}{\beta}\sum_{\omega_{n}}G^{\mathrm{HF}}(i\omega_{n})e^{i\omega_{n}0^{+}}=\quad\includegraphics[height=0.1\textwidth]{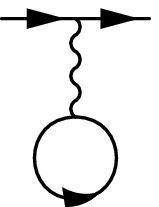}\\
\intertext{and}\mathcal{S}^{\mathrm{HF}} & =\frac{U}{\beta}\sum_{\omega_{n}}\mathcal{G}^{\mathrm{HF}}(i\omega_{n})=\quad\includegraphics[height=0.1\textwidth]{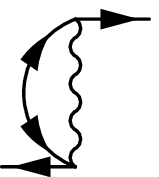},\label{eq:HFse}
\end{align}
where the HF Green function reads
\begin{widetext} 
\begin{gather}
\widehat{G}^\mathrm{HF}(i\omega_{n})=
-\frac{1}{D^\mathrm{HF}(i\omega_{n})}\begin{pmatrix}i\omega_{n}[1+s(i\omega_{n})]+\varepsilon+\Sigma^\mathrm{HF}, & -\Delta_{\Phi}(i\omega_{n})+\mathcal{S}^\mathrm{HF}\\[0.3em]
-\Delta_{\Phi}^{*}(i\omega_{n})+{\mathcal{S}^\mathrm{HF}}^{*}, & i\omega_{n}[1+s(i\omega_{n})]-\varepsilon-\Sigma^\mathrm{HF}
\end{pmatrix},\text{with the determinant}\label{eq:GFHF}
\end{gather}
\begin{gather}
D^\mathrm{HF}(i\omega_{n}) =\omega_{n}^{2}\left[1+s(i\omega_{n})\right]^{2}+\left|\varepsilon+\Sigma^\mathrm{HF}\right|^{2}+\left|\Delta_{\Phi}(i\omega_{n})-\mathcal{S}^\mathrm{HF}\right|^{2}.
\label{eq:DetHF}
\end{gather}
\end{widetext}
The hole (asterisks) functions are obtained from complex conjugation
of the equations for the electron functions. Obviously, the frequency-independent
self-energy of the HF approximation neglects any dynamical correlations
caused by particle interaction. Nevertheless, it is still capable
to describe qualitatively the $0-\pi$ quantum phase transition even
without the necessity of the common, yet questionable,\cite{Rozhkov99,Vecino03,Rodero12}
breaking of spin-reflection symmetry.\cite{Zonda15,Janis15} Therefore,
it is a useful demonstration tool of the basic ideas concerning the
model as well as a worthy etalon of more elaborate methods. The
explicit formula for the HF phase boundary for the completely symmetric
case $\Delta_{L}=\Delta_{R}$, $\Gamma_{L}=\Gamma_{R}$ was derived
in our previous work \cite{Zonda15}; here we provide a general solution. 

Before that, though, we discuss charge-conservation and gauge-invariance
properties of the HF approximation. If we insert Eq.~\eqref{eq:HFse},
into the last line of Eq.~\eqref{eq:charge-cons} we see that it
is satisfied. Thus, the HF self-energy yields a charge-conserving
approximation. Similarly, the HF self-energy transforms under the
gauge transformation identically to the (dressed) Green function as
it should. Furthermore, HF self-energy can be derived from a manifestly
gauge-invariant Luttinger-Ward functional 
\begin{multline}
\phi^{\mathrm{HF}}[\widehat{G}]=\frac{U}{\beta^{2}}\sum_{\omega_{n},\omega_{k}}\left\{ e^{i(\omega_{n}-\omega_{k})0^{+}}G(i\omega_{n})G^{*}(i\omega_{k})\right.\\
\left.\phantom{\frac{1}{2}}+\mathcal{G}(i\omega_{n})\mathcal{G}^{*}(i\omega_{k})\right\} \\
=\includegraphics[height=0.11\textwidth]{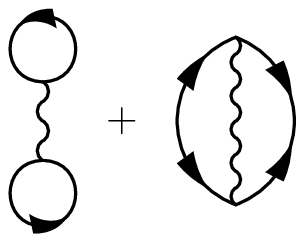}\,.\label{eq:HFLW}
\end{multline}
Consequently, HF approximation is both charge conserving as well as
thermodynamically consistent. 
\end{subequations}

Now, we turn to the calculation of the HF phase boundary. The self-consistent
HF equations read as

\begin{multline}
\begin{split}\Sigma^{\mathrm{HF}} & =-\frac{U}{\beta}\sum_{\omega_{n}}e^{i\omega_{n}0^{+}}\frac{i\omega_{n}[1+s(i\omega_{n})]+\Sigma^{\mathrm{HF}}+\varepsilon}{D^{\mathrm{HF}}(i\omega_{n})},\\
\mathcal{S}^{\mathrm{HF}} & =-\frac{U}{\beta}\sum_{\omega_{n}}\frac{\mathcal{S}^{\mathrm{HF}}-\Delta_{\Phi}(i\omega_{n})}{D^{\mathrm{HF}}(i\omega_{n})}.
\end{split}
\label{eq:HFinit}
\end{multline}
Further manipulations of
the above equations stated in Appendix \ref{App:Hartree-Fock} lead
to the following set of equations at zero temperature (since we are
interested in the phase boundary) for the auxiliary quantities $E_{d}\equiv\Sigma^{\mathrm{HF}}+\varepsilon$
and $\delta\equiv\sum_{\alpha=L,R}\Gamma_{\alpha}e^{i\Phi_{\alpha}}-\mathcal{S}^{\mathrm{HF}}$

\begin{align}
\begin{split}E_{d} & =\varepsilon+\frac{U}{2}-U\int_{-\infty}^{\infty}\frac{d\omega}{2\pi}\frac{E_{d}}{D^{\mathrm{HF}}(i\omega)},\\
\delta & =\sum_{\alpha}\Gamma_{\alpha}e^{i\Phi_{\alpha}}\\
 & -U\int_{-\infty}^{\infty}\frac{d\omega}{2\pi}\frac{\delta+\sum\limits _{\alpha}\Gamma_{\alpha}e^{i\Phi_{\alpha}}\left(\frac{\Delta_{\alpha}}{\sqrt{\Delta_{\alpha}^{2}+\omega^{2}}}-1\right)}{D^{\mathrm{HF}}(i\omega)}.
\end{split}
\label{eq:HFreform}
\end{align}

Close to the QPT the inverse denominator $1/D^{\mathrm{HF}}(i\omega)$
is dominated by its zeros at the ABS energies $\pm\omega_{0}$ (i.e.,
$D^{\mathrm{HF}}(\pm\omega_{0})=0$) which become zero at the QPT.
Therefore, we may use the expansion of the determinant to the lowest
(second) order in $\omega$ reading\footnote{See Appendix \ref{App:Hartree-Fock} for a more detailed discussion.}
$D^{\mathrm{HF}}(i\omega)\approx E_{d}^{2}+|\delta|^{2}-\left[1+\sum_{\alpha}\Gamma_{\alpha}/\Delta_{\alpha}\right]^{2}(i\omega)^{2}$
which gives us for the ABS energies close to the QPT $\omega_{0}\approx\sqrt{E_{d}^{2}+|\delta|^{2}}/(1+\sum_{\alpha}\Gamma_{\alpha}/\Delta_{\alpha})$.
Obviously, the position of QPT coincides with the situation where
$E_{d}=\delta=0$. Close to the transition the integrals are strongly
dominated by the poles at $\pm\omega_{0}$ and we can approximately
evaluate the first two leading contributions in inverse ABS energy
(first of the order $1/\omega_{0}$ while the second of
the order $1$; all other terms are at least of order $\omega_{0}$
and, thus, irrelevant at the transition) as follows: 
\begin{widetext}
\begin{subequations}
\begin{align}
\int_{-\infty}^{\infty}\frac{d\omega}{2\pi}\frac{1}{D^{\mathrm{HF}}(i\omega)} & \approx\frac{1}{2(1+\sum_{\alpha}\Gamma_{\alpha}/\Delta_{\alpha})^{2}}\int_{-\infty}^{\infty}\frac{d\omega}{\pi}\frac{1}{(\omega_{0}^{2}+\omega^{2})}=\frac{1}{2\omega_{0}(1+\sum_{\alpha}\Gamma_{\alpha}/\Delta_{\alpha})^{2}},\\
\int_{-\infty}^{\infty}\frac{d\omega}{2\pi}\frac{f(\omega^{2})}{D^{\mathrm{HF}}(i\omega)} & \approx\int_{0}^{\infty}\frac{d\omega}{\pi}\left.\frac{f(\omega^{2})}{D^{\mathrm{HF}}(i\omega)}\right|_{\omega_{0}=E_{d}=\delta=0}
\end{align}
\end{subequations}
for a smooth function $f(x)$ vanishing at zero, i.e., $f(x\to0)=0$.
Using these approximations, we finally arrive at
\begin{equation}
\begin{split}
E_{d}\left[1+\frac{U}{2\sqrt{E_{d}^{2}+|\delta|^{2}}(1+\sum_{\alpha}\Gamma_{\alpha}/\Delta_{\alpha})}\right] & =  \varepsilon+\frac{U}{2},\label{eq:HF}\\
\delta\left[1+\frac{U}{2\sqrt{E_{d}^{2}+|\delta|^{2}}(1+\sum_{\alpha}\Gamma_{\alpha}/\Delta_{\alpha})}\right] & =  \sum_{\alpha}\Gamma_{\alpha}e^{i\Phi_{\alpha}}+U\mathcal{B},
\end{split}
\end{equation}
with $\mathcal{B}$ representing the band contribution

\begin{equation}
\mathcal{B}=\intop_{0}^{\infty}\frac{d\omega}{\pi}\frac{\sum\limits _{\alpha}\Gamma_{\alpha}e^{i\Phi_{\alpha}}\left(1-\frac{\Delta_{\alpha}}{\sqrt{\Delta_{\alpha}^{2}+\omega^{2}}}\right)}{\omega^{2}\left[1+\sum_{\alpha}\frac{\Gamma_{\alpha}}{\sqrt{\Delta_{\alpha}^{2}+\omega^{2}}}\right]^{2}+\left|\sum_{\alpha}\Gamma_{\alpha}e^{i\Phi_{\alpha}}\left(\frac{\Delta_{\alpha}}{\sqrt{\Delta_{\alpha}^{2}+\omega^{2}}}-1\right)\right|^{2}}.\label{eq:Band}
\end{equation}
In Appendix \ref{App:Hartree-Fock}, we discuss the formula \eqref{eq:Band}
in more detail in the symmetric case $\Delta_{L}=\Delta_{R}=\Delta$.

To obtain the phase boundary, we sum up  squares of the two
equations~\eqref{eq:HF}:
\begin{equation}
\left(E_{d}^{2}+\left|\delta\right|^{2}\right)\left[1+\frac{U}{2\sqrt{E_{d}^{2}+|\delta|^{2}}(1+\sum_{\alpha}\Gamma_{\alpha}/\Delta_{\alpha})}\right]^{2}  =  \left(\varepsilon+\frac{U}{2}\right)^{2}+\left|\sum_{\alpha=L,R}\Gamma_{\alpha}e^{i\Phi_{\alpha}}+U\mathcal{B}\right|^{2},\label{eq:close-to-boundary}
\end{equation}
which yields at the phase boundary $E_{d}^{2}+|\delta|^{2}=0$ an
implicit equation for the borderline
\begin{equation}
\frac{U^{2}}{4(1+\sum_{\alpha}\Gamma_{\alpha}/\Delta_{\alpha})^{2}}=\left(\varepsilon+\frac{U}{2}\right)^{2}+\left|\sum_{\alpha=L,R}\Gamma_{\alpha}e^{i\Phi_{\alpha}}+U\mathcal{B}\right|^{2}.\label{eq:phase-boundary}
\end{equation}
Omitting the band contribution $\mathcal{B}$ from Eq.~\eqref{eq:phase-boundary}
one gets a very simple approximation of the boundary called \textit{generalized
atomic limit} (GAL)\cite{Zonda15}: 
\begin{equation}
\frac{U^{2}}{4(1+\sum_{\alpha}\Gamma_{\alpha}/\Delta_{\alpha})^{2}}=\left(\varepsilon+\frac{U}{2}\right)^{2}+\Gamma_{L}^{2}+\Gamma_{R}^{2}+2\Gamma_{L}\Gamma_{R}\cos(\Phi_{L}-\Phi_{R}).\label{eq:GAL}
\end{equation}
\end{widetext}
Interestingly, as we show later on, at half-filling $(\varepsilon=-U/2)$
the simple GAL boundary lies typically very close to the results obtained
via the NRG method and/or the second-order perturbation expansion.  
This is not surprising, since at half-filling the HF approximation reproduces the atomic limit exactly. 
Hence, one can expect that GAL, being a generalization of the atomic limit to non-integer occupation of the dot, 
delivers quite reliable results near the charge-symmetric state.  

Furthermore, we may use Eq.~\eqref{eq:close-to-boundary} for finding
$\sqrt{E_{d}^{2}+|\delta|^{2}}$ close to the phase boundary. Using
the implicit-function theorem we see that a solution with $\sqrt{E_{d}^{2}+|\delta|^{2}}>0$
exists on one side of the boundary (moreover, the side containing
the noninteracting $U\to0$ limit, i.e., corresponding to the $0$ phase)
while $\sqrt{E_{d}^{2}+|\delta|^{2}}<0$ on the other side. There,
we must conclude, no solution to the restricted HF equations \eqref{eq:HF}
exists, only when one allows for breaking of the spin symmetry
(i.e., finite magnetization), which is however unphysical for a zero-dimensional
impurity system, the appropriately extended HF equations \eqref{eq:HFinit}
do have a solution.\cite{Vecino03,Rodero12} Since we do not want
to resort to an unphysical symmetry breaking to obtain the $\pi$ phase,
we must conclude that the perturbative spin-symmetric solution breaks
down at the phase boundary as expected from a general conceptual viewpoint.\cite{Anderson97}
Although these findings have been explicitly demonstrated on the level
of the HF approximation, they are actually fully general and apply
to any order of perturbation theory, in particular also to the second
order which we are going to address now.

\subsection{Second order: Dynamical corrections}

It was shown in previous studies\cite{Zonda15,Janis15} that
inclusion of dynamical corrections beyond the static HF into the self-energy
can dramatically improve the quantitative predictions for both the
position of the phase boundary and the physical quantities in the
$0$ phase. Already, first corrections from second order of the perturbation
expansion were sufficient to reproduce fairly well the results of
NRG in the case of identical leads. Second-order contributions to
the self-energy read as
\begin{widetext}
\begin{subequations}\label{eq:2ndse}
\begin{align}
\Sigma^{(2)}(i\omega_{n}) & =-\frac{U^{2}}{\beta}\sum_{\nu_{m}}G(i\omega_{n}+i\nu_{m})\chi(i\nu_{m})=\quad\includegraphics[height=0.1\textwidth]{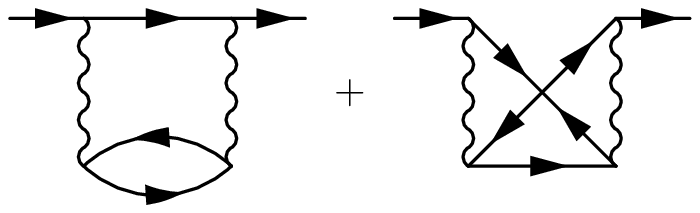}\\
\intertext{and}\mathcal{S}^{(2)}(i\omega_{n}) & =-\frac{U^{2}}{\beta}\sum_{\nu_{m}}\mathcal{G}(i\omega_{n}+i\nu_{m})\chi(i\nu_{m})=\quad\includegraphics[height=0.1\textwidth]{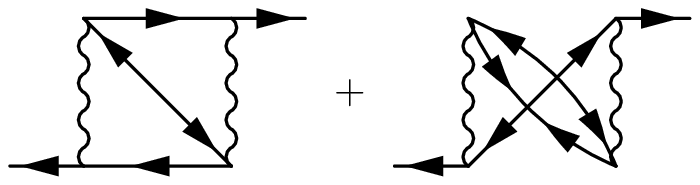}\label{eq:2ndsean}
\end{align}
where 
\begin{gather}
\begin{split}\chi(i\nu_{m}) & =\frac{1}{\beta}\sum_{\omega_{k}}\left[G(i\omega_{k})G^{*}(-i\nu_{m}-i\omega_{k})+\mathcal{G}(i\omega_{k})\mathcal{G}^{*}(-i\nu_{m}-i\omega_{k})\right]\\
 & =\frac{1}{\beta}\sum_{\omega_{k}}\left[G(i\omega_{k})G(i\nu_{m}+i\omega_{k})+\mathcal{G}(i\nu_{m}+i\omega_{k})\mathcal{G}^{*}(i\omega_{k})\right]
  =\chi(-i\nu_{m})=\chi^{*}(i\nu_{m})
\end{split}
\label{eq:bubble}
\end{gather}
is the two-particle bubble consisting of the normal and anomalous
parts and $\nu_{m}=2\pi m/\beta$ is the $m$-th bosonic Matsubara
frequency. Analogously to the HF case before we can explicitly verify
the charge conservation condition \eqref{eq:charge-cons}; calculations
are more tedious this time and we present them in Appendix ~\ref{App:Charge-conservation}.

The Luttinger-Ward functional of this second-order correction to the
Hartree-Fock approximation reads as
\begin{equation}
\begin{split}
\phi^{(2)}[\widehat{G}]&=-\frac{U^{2}}{2\beta^{3}}\sum_{\omega_{n},\omega_{k},\nu_{m}}\left[G(i\omega_{k})G^{*}(-i\omega_{n})G(i\omega_{n}+i\nu_{m})G^{*}(-i\omega_{k}-i\nu_{m})+2G(i\omega_{k})G^{*}(-i\omega_{n})\right.\\
&\qquad\quad\left.\times\mathcal{G}(i\omega_{n}+i\nu_{m})\mathcal{G}^{*}(-i\omega_{k}-i\nu_{m})+\mathcal{G}(i\omega_{k})\mathcal{G}^{*}(-i\omega_{n})\mathcal{G}(i\omega_{n}+i\nu_{m})\mathcal{G}^{*}(-i\omega_{k}-i\nu_{m})\right]\\
&=\includegraphics[width=0.4\textwidth]{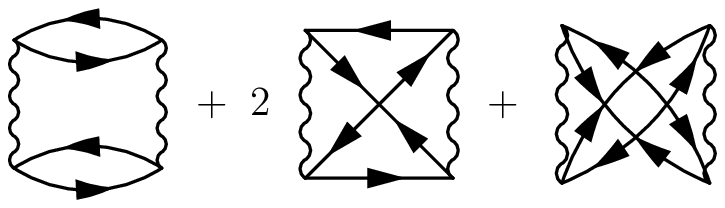}\ .
\end{split}
\end{equation}
\end{subequations}
\end{widetext}
It is manifestly gauge-invariant. These first two orders of the
perturbation expansion are well controllable on the one-particle level.
The higher contributions to the self-energy become more complex and
their classification demands to introduce two-particle vertices as discussed in detail in Ref.~\onlinecite{Janis14}.
Therefore, we resort just to the second order of the perturbation
theory, which proves to be fully sufficient in the BCS-singlet
regime for weak and intermediate coupling. The second order self-energy corrections (together with the first-order HF counterparts) are inserted into Eq.~\eqref{eq:GF}
to obtain a self-consistent nonlinear functional equation for the
Green function as a function of frequency. Unlike the HF case,
the resulting equations for the Green function components defy analytical
treatment and must be solved numerically. In the following we refer
to this approach as the \textit{full self-consistent dynamical
correction} (FDC) approximation. 

As discussed previously for the symmetric leads \cite{Zonda15} nearly identical results can be obtained in the weak coupling regime 
by evaluating the dynamical self-energies \eqref{eq:2ndse} using just
the fully converged self-consistent HF solution as the input into
the Green function (DC approximation). 
The DC approach can be represented by the following algorithm:
\begin{enumerate}
  \item Compute the HF Green function as described in Sec.~\ref{HFA}.  
  \item Compute the second-order contributions to self energy ${\Sigma^{(2)}}(i\omega_{n})$ 
   and ${\mathcal{S}^{(2)}}(i\omega_{n})$ using formulas \eqref{eq:2ndse} 
   with  $G^\mathrm{HF}(i\omega_{n})$ and $\mathcal{G}^\mathrm{HF}(i\omega_{n})$ instead of $G(i\omega_{n})$ and $\mathcal{G}(i\omega_{n})$. 
   These second-order contributions stay fixed  throughout further calculations.
  \item Compute the DC self-energies ${\Sigma}^\mathrm{DC}(i\omega_{n})={\Sigma^{(1)}}+{\Sigma^{(2)}}(i\omega_{n})$ and 
${\mathcal{S}}^\mathrm{DC}(i\omega_{n})={\mathcal{S}^{(1)}}+{\mathcal{S}^{(2)}}(i\omega_{n})$ with ${\Sigma^{(1)}}={\Sigma}^\mathrm{HF}$ and 
${\mathcal{S}^{(1)}}={\mathcal{S}}^\mathrm{HF}$ in the first iteration. 
  \item Compute the DC Green function $\widehat{G}^\mathrm{DC}(i\omega_{n})$ using definitions \eqref{eq:GF} and \eqref{eq:Det} with 
   ${\Sigma}(i\omega_{n})={\Sigma}^\mathrm{DC}(i\omega_{n})$ 
   and ${\mathcal{S}}(i\omega_{n})={\mathcal{S}}^\mathrm{DC}(i\omega_{n})$.
  \item Compute the first order contributions to self energies: 
${\Sigma^{(1)}}=\frac{U}{\beta}\sum_{\omega_{n}}G^\mathrm{DC}(i\omega_{n})e^{i\omega_{n}0^{+}}$
and ${\mathcal{S}^{(1)}}=\frac{U}{\beta}\sum_{\omega_{n}}\mathcal{G}^\mathrm{DC}(i\omega_{n})$.
  \item Repeat steps 3 to 5 until the convergence criterion of the self-consistency is achieved.
\end{enumerate}
The algorithm implies that the convolutions in the second-order
self-energies are evaluated just at the beginning of the procedure. 
The fixed dynamical self-energies are then used to calculate self-consistently the first-order contributions to the self-energies. 

Note that the DC approximation is numerically more stable close to the phase transition than the FDC approximation (see Fig.~\ref{fig:ABS}).  In addition, it allows us to study the intermediate coupling regimes, where full self-consistent approaches often fail to give physically correct solution. It is fairly well known that the fully self-consistent second-order approximation as well as its extensions via sums of ladders and chains  fail for intermediate coupling of impurity models. They not only smear the Hubbard satellite bands\cite{Janis07}, they also miss the Kondo physics.\cite{Hamann:1969aa} That is why simplified self-consistencies often provide better approximations than the fully self-consistent ones.\cite{Georges:1996aa}

\section{Results and Discussion\label{sec:Results}}

We provide a comparison of the ground-state, i.e. zero-temperature,
results obtained via the perturbative method discussed above with
those obtained using the NRG approach which is a reliable nonperturbative
numerical method for the ground state properties.\cite{Bulla08} For
NRG calculations we used the NRG Ljubljana open source code\cite{NRGLjubljana,Zitko09}
mostly with the logarithmic discretization parameter $\Lambda$ set
to the value of $4$ as is common for double-channel problems. 
We also tested and used other values of $\Lambda$ (see Fig.~\ref{fig:GammaEpsU}a)
and found out that for most of the studied cases, the phase boundary
is not sensitive to $\Lambda$. This is in compliance with the findings
discussed in Ref.~\onlinecite{Tanaka07}.

\begin{figure}
\includegraphics[width=1\columnwidth]{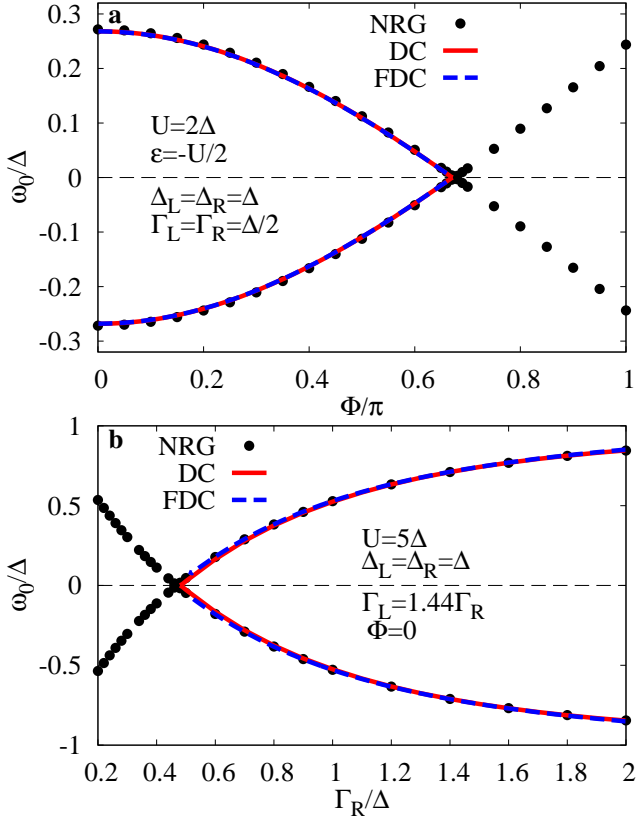}\caption{(Color online) Andreev bound states energies dependence on the phase
difference $\Phi$ for the symmetric coupling $\Gamma_{L}=\Gamma_{R}$
\textbf{(a)} and on the coupling $\Gamma_{R}$
for asymmetric coupling $\Gamma_{L}=1.44\Gamma_{R}$ at $\Phi=0$
\textbf{(b)}. The solid lines were calculated via
the DC approximation, the dashed lines via the FDC approximation and the bullets were obtained using NRG with
$\Lambda=4$. The point where bound and anti-bonding states merge/cross
at the Fermi level (zero energy) is identified with the $0-\pi$ quantum
phase transition. \label{fig:ABS}}
\end{figure}
Various theoretical studies showed that the $0-\pi$ phase transition,
where the supercurrent changes its sign, is accompanied by a smooth
crossing of the Andreev bound states\cite{Clerk00,Choi04,Oguri04,Luitz10,Pillet13}
which is in agreement with the experiments.\cite{Pillet10,Pillet13,Jesper13}
Although the perturbation approach without spin-symmetry breaking
can not be easily extended into the $\pi$ phase and, therefore, does
not show the actual crossing of the ABS,\cite{Zonda15} the ABS smoothly
reach the Fermi energy at the border of the $0$-phase.\cite{Zonda15,Janis15}
In Fig.~\ref{fig:ABS} we plot examples of the ABS dependencies on
the phase difference $\Phi$ for symmetric coupling $\Gamma_{L}=\Gamma_{R}$
(Fig.~\ref{fig:ABS}a) as well as on the right coupling $\Gamma_{R}$
for asymmetric coupling $\Gamma_{L}=1.44\Gamma_{R}$ (Fig.~\ref{fig:ABS}b).
We identify the point where both (bound and anti-bonding) ABS reach
the Fermi energy with the boundary of the $0$-phase, i.e., with the
point of the quantum phase transition. This is fully supported by
the NRG data. It can be seen in Fig.~\ref{fig:ABS} that the crossing
of the ABS obtained by the NRG (bullets) coincides with the merger
of the ABS obtained via the DC approximation (solid red lines).
The dashed blue lines in Fig.~\ref{fig:ABS} were obtained via
the FDC approximation. One can see that the FDC and DC
results practically coincide in both presented cases apart from the very close neighborhood of the phase transition, where the FDC becomes numerically unstable.         

It should be stressed that analytic continuation of the Matsubara
formalism to the real frequencies is necessary for the study of ABS.
This usually leads to quite complicated formulas for the Green functions,\cite{Janis15}
however, sometimes this approach is numerically more stable than the
Matsubara formalism. Nevertheless, the continuation to the real axis
can be avoided if one is interested only in the phase boundaries.
The determinant $D(i\omega_{n})$ yields zero at the phase transition
point exactly for $i\omega_{n}=0$. Therefore, one can use the smooth
dependency of $D(0)$ on different model parameters for the direct
estimation of the phase boundary.

\subsection{Phase diagrams for $\Delta_{L}=\Delta_{R}$}

Setups where $\Delta_{L}=\Delta_{R}\equiv\Delta$ in practice mean
that both leads are made of the same material. Since this is a common
situation in the experiment, this case has been intensively studied.\cite{Yeyati11}
We start our discussion with three phase diagrams known from the literature.\cite{Tanaka07}
The main reason for this is to show that the simple second-order expansion
theory is sufficient for a broad range of parameters and even in the
regimes where usually much more elaborated techniques are used. Simultaneously,
we compare the results of the three approximations, namely, HF, DC, and
GAL and discuss their limitations.

\begin{figure}
\includegraphics[width=0.95\columnwidth]{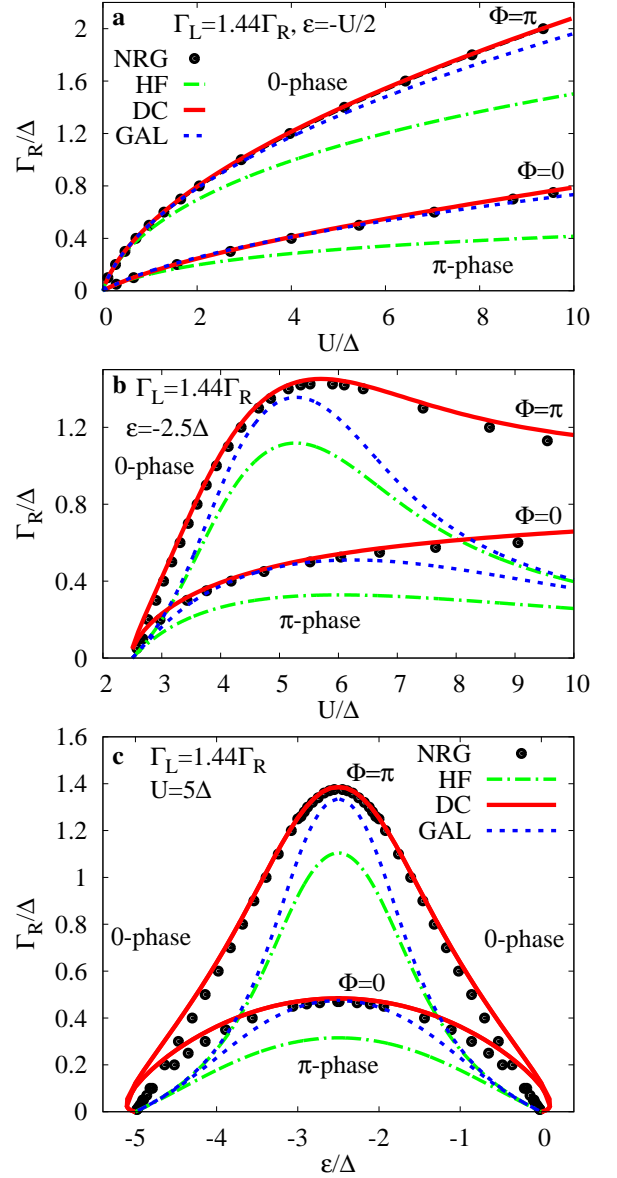}
\caption{(Color online) Phase diagrams in the $\Gamma_{R}-U$ plane at the
half-filling (\textbf{a}) and for $\varepsilon=-2.5\Delta$ (\textbf{b})
and in the $\Gamma-\varepsilon$ plane for $U=5\Delta$ (\textbf{c}).
We compare the phase boundaries calculated by NRG with the spin-symmetric
HF, the second-order PT/dynamical corrections (DC), and generalized
atomic limit approximation (GAL). \label{fig:Phase-diagrams1}}
\end{figure}
In Fig.~\ref{fig:Phase-diagrams1}a, we plot the ground-state phase
diagrams in the $\Gamma_{R}-U$ plane for $\Phi=0$ and $\Phi=\pi$
at half-filling ($\varepsilon=-U/2$). The left-right lead asymmetry
is fixed by the coupling ratio $\Gamma_{L}=1.44\Gamma_{R}$. The first
thing that should be noticed is that the HF approximation without
broken spin-symmetry yields a qualitatively correct description of
the phase boundary at half-filling. However, its phase-boundary curve
is close to the NRG border only for small Coulomb interaction ($U\lesssim2\Delta$).
A much better result is obtained by using its simplification, namely,
the GAL approximation, which neglects the band contribution. This
implies that the HF approximation overestimates the contribution from
the bands. Considering its simplicity, the GAL method provides a very
good and fast approximation of the phase boundary at half-filling
even for $U\gg\Delta$. Nevertheless, regarding the accuracy it is
overperformed by the DC approximation. One can see that the DC border
reproduces the NRG data (for $\Lambda=4$) almost perfectly. This
shows that the perturbation theory with the simplest dynamical corrections
can lead to a correct estimation of the quantum phase boundary for
the studied system. This statement is true not only for the symmetric-leads
case studied in Ref.~\onlinecite{Zonda15}, but also for the experimentally
more relevant setups.

A similar agreement between the DC and NRG phase boundaries can be
seen in Fig.~\ref{fig:Phase-diagrams1}b, where the phase diagrams
are plotted away from half-filling at $\varepsilon=-2.5\Delta$. On
the other hand, the bias of the level energy $\varepsilon$ strongly
influences both the HF and GAL curves. The GAL border approaches the
NRG data only around $U=5\Delta$, i.e., just near the half-filling
occupation. The HF border is way off in the whole plotted range. In
addition, both HF and GAL results drift away from the DC and NRG border
with the increasing $U$. Because of the structure of Eq.~\eqref{eq:phase-boundary},
the GAL and HF borders approach each other for $U\rightarrow0$, where
the term $U\mathcal{B}$ vanishes, as well as for $|\epsilon+U/2|\gg\Delta$
where the first term dominates the right hand side of Eq.~\eqref{eq:phase-boundary}.
The latter one can be seen in Fig.~\ref{fig:Phase-diagrams1}c. Here,
we plot the phase diagram in the $\Gamma_{R}-\varepsilon$ plane for
moderately large Coulomb interaction $U=5\Delta$ and two values of
$\Phi$. The HF and GAL borders coincide far away from half-filling
for both values of $\Phi$. As before, despite its simplicity, both
these approximations yield a fair qualitative agreement with the NRG
data. However, with the exception of the GAL curve near the half-filling,
both approximations fail to reproduce the NRG data quantitatively.
In contrast, the DC border matches the NRG in a broad region around
$\varepsilon=-U/2$ and even outside this region the difference between
NRG data and DC curve is much smaller than $\Delta$. This shows that
the proper treatment of the frequency dependence of the correlation
effects is crucial for the quantitative description of the $0-\pi$
transition.

\begin{figure}
\includegraphics[width=0.95\columnwidth]{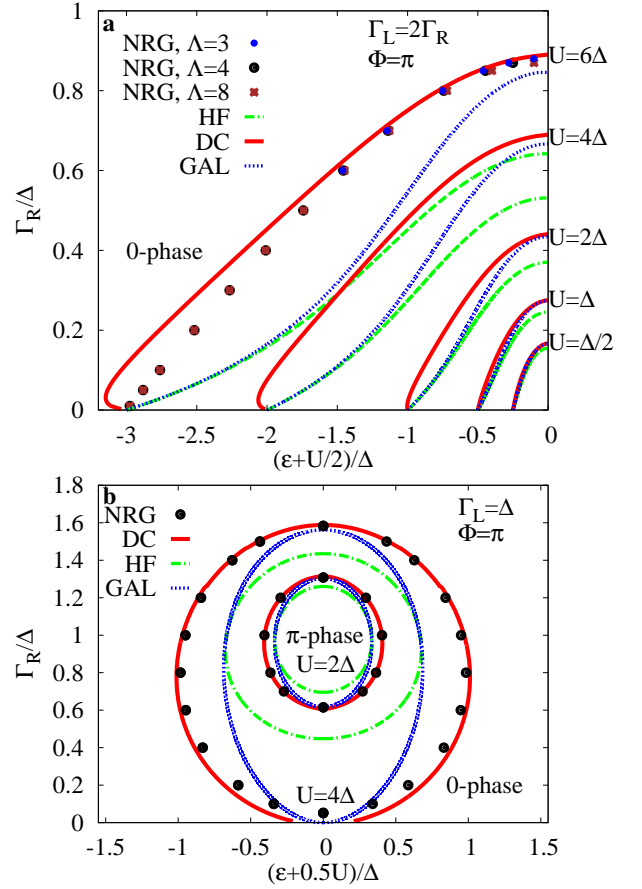}
\caption{(Color online) Phase diagrams in the $\Gamma_{R}-\varepsilon$ plane
for different values of $U$, $\Phi=\pi$ and asymmetric leads $\Gamma_{L}=2\Gamma_{R}$
(fixed ratio) \textbf{(a)} and $\Gamma_{L}=\Delta$
(varying ratio $\Gamma_{R}/\Gamma_{L}$ which induces the $\pi$-phase
island structure) \textbf{(b)}.\label{fig:GammaEpsU}}
\end{figure}

The most evident (qualitative) discrepancy of the DC curve from the
NRG are the ``humps'' at the bottom of the phase diagrams Figs.~\ref{fig:Phase-diagrams1}c and \ref{fig:GammaEpsU}a. 
This is not surprising as here $U\gg\Gamma$ and therefore we are on the edge
of the usability of the perturbation expansion in $U$ (see Fig.~\ref{fig:Phase-diagram-Intro}).
On the other hand these humps resemble a formation of the island-like
phase diagram known from the previous NRG study by Oguri \emph{et
al.}\cite{Oguri04}. They showed that in case of strongly asymmetric
gaps $\Delta_{L}\gg\Delta_{R}$ and decreasing $U$ a re-entrant doublet
region appears as an island in the $\Gamma_{R}-\varepsilon$ plane
(see Fig.~12 in Ref.~\onlinecite{Oguri04}). However, this is not the
case in Fig.~\ref{fig:GammaEpsU}a. For the fixed ratio $\Gamma_{L}/\Gamma_{R}$
and $\Delta_{L}=\Delta_{R}$, the humps are only a consequence of the
perturbative treatment. Unlike in Ref.~\onlinecite{Oguri04} they are obviously
disappearing with the decreasing $U$ (Fig.~\ref{fig:GammaEpsU}a.)
and no island structure appears even for $U=\Delta/2$. Nevertheless, the
situation is different if one varies the ratio $\Gamma_{L}/\Gamma_{R}$.
In Fig.~\ref{fig:GammaEpsU}b we show that the condition $\Delta_{L}\gg\Delta_{R}$
used by Oguri \emph{et al.} is not necessary for obtaining the island structures
of the $\pi$-phase. All three approximations show these structures
when the ratio $\Gamma_{L}/\Gamma_{R}$ is varied. Nevertheless, as
before, only the DC approximation matches quantitatively with the
NRG outside the half-filling for plotted values of $U$.

Because we have observed this behavior also for other values of $\Gamma_{L}/\Delta$
(for another example see the reentrant behavior at $\Delta_{R}/\Delta_{L}=1$
for $U=\Delta_{L}$ lines in Fig.~\ref{fig:PhD_DLGL}a), we conclude
that the island structures are primarily a function of the difference
between the hybridizations $\Gamma_{L}$ and $\Gamma_{R}$. In general,
the $\pi$ phase is destabilized by the increasing differences between
the couplings. As this can be in principle tuned in the experiment\cite{DeFranceschi10}
one can expect that the island structures can be verified experimentally.
Regarding the $U$ dependence, one can see that the increasing Coulomb
interaction inflates the $\pi$-phase area in both panels of Fig.~\ref{fig:GammaEpsU}. 

We use Fig.~\ref{fig:GammaEpsU}a to demonstrate two technicalities.
First, it is an illustrative way to show that all three approximations
converge to each other with decreasing $U$ as it should be. All borders
practically coincide for the lowest $U=\Delta/2$. Second, we present
a test of the NRG for the largest value of Coulomb interaction $U=6\Delta$
corresponding to the most correlated case. There is only a very weak
dependence of the NRG phase boundary on the parameter $\Lambda$ which
makes the generic calculations at $\Lambda=4$ highly reliable.

\subsection{Phase diagrams for $\Delta_{L}\protect\neq\Delta_{R}$}

Significantly less attention has been paid, both experimentally and theoretically,
to the general case $\Delta_{L}\neq\Delta_{R}$ than to the identical leads $\Delta_{L}=\Delta_{R}$. This can change
in the near future as a recent experiment on a carbon nanotube quantum
dot coupled to Nb fork and Al tunnel probe\cite{Kumar14} not only
showed that such a setup is technically possible, but in addition it
revealed a nontrivial formation of ABS. Theoretical
efforts so far have been, however, mainly focused on the $\Delta_{L}=\Delta_{R}$
case despite the fact that previous theoretical studies dealing with
some special cases of non-identical leads showed that the lead difference
can strongly affect the quantum phase transition.\cite{Oguri04,Tanaka07}

\begin{figure}
\includegraphics[width=0.95\columnwidth]{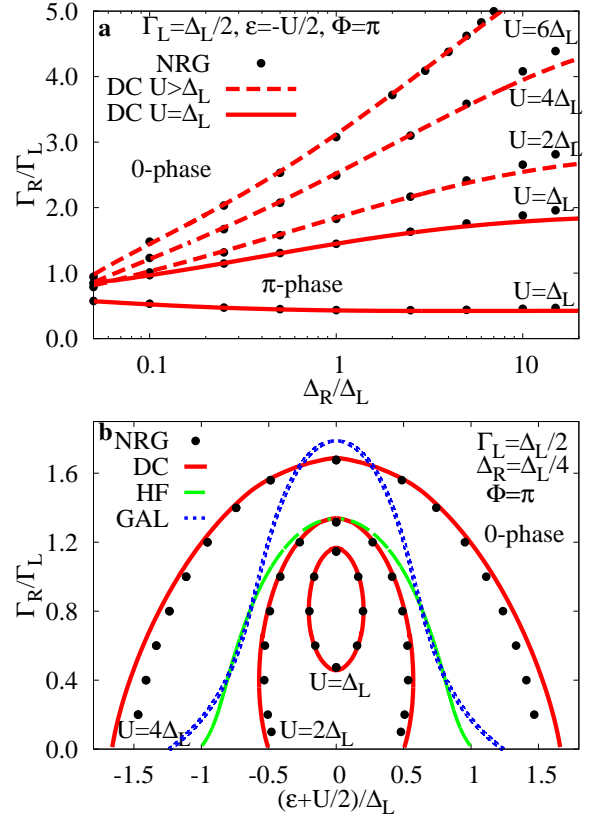}
\caption{(Color online) a) Critical tunnel-coupling ratio $\Gamma_{R}/\Gamma_{L}$
as a function of SC-gap ratio $\Delta_{R}/\Delta_{L}$ for $\Gamma_{L}=\Delta_{L}/2$,
$\Phi=\pi$ at half-filling. b) Phase diagrams in the $\Gamma_{R}/\Gamma_{L}-\varepsilon$
plane with asymmetric lead SC gaps $\Delta_{R}=\Delta_{L}/4$ calculated
via NRG and DC approximation for $U/\Delta_{L}=1,2,4$ and compared
for $U=4\Delta_{L}$ with HF and GAL approximations. \label{fig:PhD_DLGL}}
\end{figure}

We demonstrate this for a wide range of ratios $\Delta_{R}/\Delta_{L}$
in Fig.~\ref{fig:PhD_DLGL}a. Here, the phase boundary obtained via
the DC approximation and NRG is plotted in the $\Gamma_{R}/\Gamma_{L}$
- $\Delta_{R}/\Delta_{L}$ plane for various values of $U$ at half-filling.
We have set the coupling to the left lead to $\Gamma_{L}=\Delta_{L}/2$
and the phase difference to $\Phi=\pi$. One can see that the DC curves
are in excellent agreement with the NRG results in the plotted range
of $\Delta_{R}/\Delta_{L}$ which spans two orders of magnitude (note
the logarithmic horizontal scale). This once again proves the reliability
of the DC approximation also for unequal gaps in the leads. 

We can observe two different kinds of phase diagrams in Fig.~\ref{fig:PhD_DLGL}a.
Two phase boundaries separating the $0$- and $\pi$-phases are present
for $U=\Delta_{L}$ (solid curves), but only a single phase boundary
is realized in the plotted region for $U\geq2\Delta_{L}$ (dashed
curves). This is in compliance with the study of $\Delta_{R}\gg\Delta_{L}$
case in Ref.~\onlinecite{Oguri04} as well as with the opening of island
structures as a consequence of increasing $U$ observed in the phase
diagrams plotted in Figs.~\ref{fig:GammaEpsU}b and \ref{fig:PhD_DLGL}b. 

Oguri \emph{et al.}\cite{Oguri04} showed that in the limit $\Delta_{R}\rightarrow\infty$
the model \eqref{eq:Ham} can be mapped onto a single-channel model
where the right lead is replaced by an onsite superconducting gap
at the impurity, i.e., the standard superconducting atomic limit\cite{Bauer07,Karrasch08}
performed for the right lead only. Our observation that the critical
$\Gamma_{R}$ depends only weakly on the ratio $\Delta_{R}/\Delta_{L}$
for the large right gap and weak Coulomb interaction (see $U=\Delta_{L}$
boundaries in Fig.~\ref{fig:PhD_DLGL}a) justifies applicability
of this simplified model.

The GAL and HF phase boundaries are in a fair qualitative agreement with DC and NRG even
for $\Delta_{L}\neq\Delta_{R}$ case. Nevertheless, both these simple
approximations fail quantitatively in positioning the critical
curves in the phase diagram. This can be seen in Fig.~\ref{fig:PhD_DLGL}b where the $0$-phase boundaries are plotted in the $\Gamma_{R}/\Gamma_{L}$ - $\varepsilon$
plane for $\Delta_{R}=\Delta_{L}/4$ and $\Gamma_{R}=\Gamma_{L}/2$.
On the other hand, the DC approximation largely coincides with NRG
even for $U=4\Delta_{L}$ and fully reproduces the $\pi$-phase island
structure obtained with NRG for $U=\Delta_{L}$.

\subsection{Applicability and limitations of the method\label{sub:Applicability}}

Despite its reliability in a wide range of the input parameters the present method has natural application limits. We can roughly divide them into two classes. The first one is related
to the very conceptual foundation of the method in the (resummed)
perturbation theory which implies its breakdown at the quantum phase
transition and impossibility to reach the $\pi$ phase with the doubly-degenerate
ground state. Any quantities in the $\pi$ phase are presently inaccessible
by our approach. The impossibility to take into account the $\pi$ phase has a serious
consequence in that the method cannot address  nonzero temperatures. At nonzero temperatures both the $0$ and $\pi$ phases coexist
and this coexistence results in a temperature-smoothened behavior of
all quantities around the transition. This is, however, completely
neglected in the present perturbative treatment that is built upon only the singlet equilibrium state, $0$ phase, even if it becomes metastable (at non-zero temperature)
and should actually be replaced by the  spin doublet, $\pi$ phase. Mixing of the singlet and doublet states  is relevant only in a close vicinity of the phase transition, since
far away from it the physics is still described well by either of the states. Yet, this
conceptual limitation is serious, especially, since there is no clear
way how to circumvent it.  There is, however, a perturbative expansion for elementary excitations and Green 
functions in systems with a degenerate equilibrium state.\cite{Brouder09, *Louie12} 

The second class of limitations concerns the standard fact that perturbation
methods have limited range of applicability given by the level of
sophistication of the included perturbation contributions. Such theories
typically do not explicitly break down but they become quantitatively
highly imprecise and eventually useless. Our approach is conceptually
meaningful in the $0$ phase and can be used not only for  determination
of the position of $0-\pi$ phase boundaries, but also for calculation
of various (single-particle) quantities such as the Josephson current,
QD occupation, proximity-induced local gap, energies of ABS etc.\cite{Zonda15,Janis15}
Since it is a perturbation expansion in the Coulomb interaction truncated
at the lowest-order diagrams, it is clear that it  ceases to be reliable  for
large enough $U$. This can happen in various ways depending on the
values of the other model parameters $\Delta$, $\Gamma$, and $\varepsilon$.
We have already encountered such a situation in Figs.~\ref{fig:Phase-diagrams1}c
and \ref{fig:GammaEpsU}a where there was an obvious discrepancy (although
not too severe) for small $\Gamma$ close to charge-degeneracy points.
Small $\Gamma$ effectively increases the importance of the Coulomb
interaction via the increased ratio $U/\Gamma$, pushing the system
close to the atomic limit, where the complementary perturbation expansion
in $\Gamma$ is a more suitable choice.\cite{Glazman89,Novotny05} 

\begin{figure}
\includegraphics[width=0.9\columnwidth]{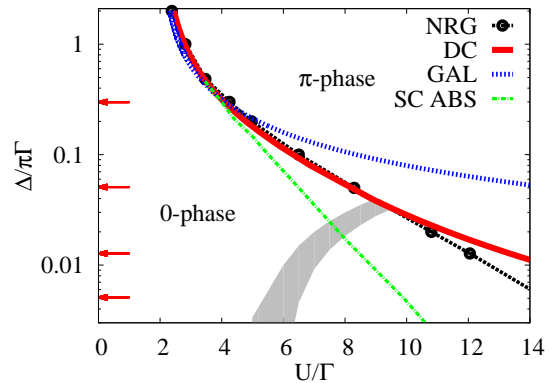}\caption{(Color online) 
Phase diagram in the $\Delta-U$ plane for $\phi=0$. NRG solution
(black bullets connected by dashed lines) is compared with the DC
approximation (red full line), GAL (blue dotted line), and the SC
ABS method taken graphically from Fig.~3 in Ref.~\onlinecite{Meng09}
(green dot-dashed line). Notice the logarithmic scale on the vertical
$\Delta$ axis, where the arrows point out the values of the gap used
for curves plotted in Fig.~\ref{fig:One-Particle-Quantities}. The
grey stripe marks the region where DC results start to depart from
the NRG in Fig.~\ref{fig:One-Particle-Quantities}, i.e., the position
of the crossover from the BCS to the Kondo singlet ground state.\label{fig:PhD_U_delta}}
\end{figure}

\begin{figure}
\includegraphics[width=0.9\columnwidth]{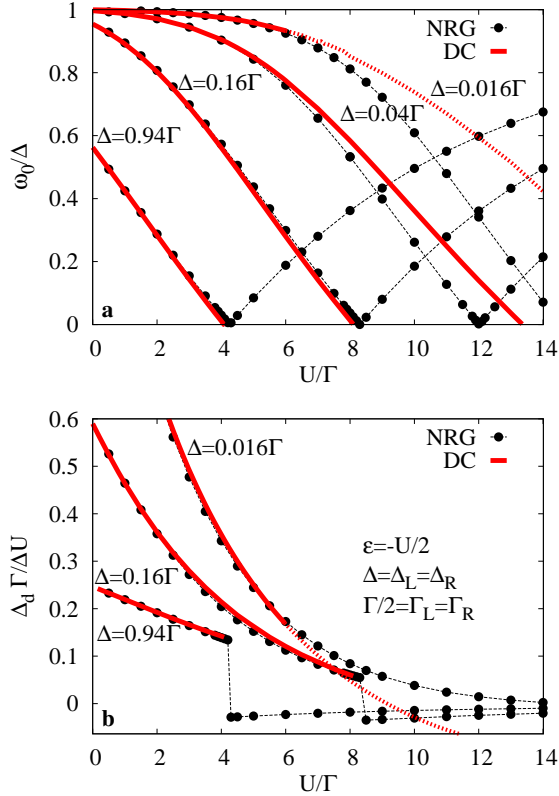}\caption{(Color online) Limitations of the DC approximation --- comparison
of the NRG (bullets) and DC approximation (solid lines) results for
symmetric setups at half-filling. Interaction strength $U$ dependence
of the ABS energies \textbf{(a)} and scaled proximity-induced
gap $\Delta_{d}=U\left\langle d_{\uparrow}d_{\downarrow}\right\rangle $
\textbf{(b)} for the four values of the gap $\Delta$
shown by arrows in Fig.~\ref{fig:PhD_U_delta} and taken from Ref.~\onlinecite{Bauer07}
(the curve for $\Delta=0.04\Gamma$ was omitted in the lower panel
just for its readability). The breakdown of the DC method can be seen
for large interaction $U/\Gamma$ and small SC gap $\Delta$ as indicated
in Fig.~\ref{fig:PhD_U_delta}. The dashed part of the $\Delta=0.016\Gamma$
lines is numerically unstable. \label{fig:One-Particle-Quantities}}
\end{figure}

For decreasing $\Gamma$ and fixed $\Delta$, the perturbation expansion
around the atomic limit works fine, but if one allows also the SC
gap $\Delta$ to decrease comparably to the (normal state) Kondo temperature
$T_{K}\sim\sqrt{\frac{U\Gamma}{2}}\exp\left[\pi\frac{\varepsilon}{2\Gamma}\left(1+\frac{\varepsilon}{U}\right)\right]$,
the system enters into a strongly correlated Kondo state where any
simple perturbation theory inevitably fails. We demonstrate this crossover
from the conventional BCS singlet to the Kondo singlet\cite{Bauer07}
and gradual failure of the DC approximation in Figs.~\ref{fig:PhD_U_delta}
and \ref{fig:One-Particle-Quantities} for symmetric leads at half-filling.
In Fig.~\ref{fig:PhD_U_delta} we present the phase diagram in the
$\Delta-U$ plane studied previously in the literature using different
methods including the NRG\cite{Bauer07} and the expansion around
the atomic limit (for $\Delta$ much larger than the characteristic energies
of the dot) based on the self-consistent description of the Andreev
bound states (SC ABS).\cite{Meng09} We compare the phase boundaries
obtained via these methods with the DC and GAL boundaries in Fig.~\ref{fig:PhD_U_delta}.
Note, that the DC approximation is so good that we had to use the
logarithmic scale for the $\Delta$-axis to visualize deviations of
the DC boundary from the NRG data. The DC boundary departs from the
NRG points for $\Delta/\pi\Gamma\lesssim0.03$, nevertheless, even
in this parameter range the DC boundary is still much closer to the
NRG boundary than the SC ABS curve, which can be attributed to the
violation of the large-$\Delta$ assumption inherent to the SC ABS
approach. However, the GAL branches off from NRG points significantly
with decreasing $\Delta$. 

The horizontal arrows in Fig.~\ref{fig:PhD_U_delta} correspond to
the values of the gap for which the one-particle quantities are plotted
in Fig.~\ref{fig:One-Particle-Quantities}a (ABS energies $\omega_{0}$)
and Fig.~\ref{fig:One-Particle-Quantities}b (proximity-induced local
gap $\Delta_{d}=U\left\langle d_{\uparrow}d_{\downarrow}\right\rangle $;
the curve $\Delta=0.04\Gamma$ is not displayed just for clarity).
There is almost a perfect agreement between DC and NRG curves for
$\Delta/\Gamma>0.1$ in the entire $0$-phase. For smaller values
of $\Delta$ and sufficiently large $U$ both ABS energies and $\Delta_{d}$
obtained using the DC approximation depart from the NRG points with
increasing ratio $U/\Gamma$ and can even become numerically unstable
(dashed lines). We estimated the region where the DC curves start
to deviate significantly from the NRG by the grey stripe in Fig.~\ref{fig:PhD_U_delta}.
We consider this to be the crossover region between the BCS and Kondo
singlet ground states and, therefore, also the edge of applicability
of the DC approximation. As we will show in the next section, this
edge still leaves plenty of space for the second-order expansion in
$U$ to be the method of choice for the description of real systems. 

\begin{figure}
\includegraphics[width=1\columnwidth]{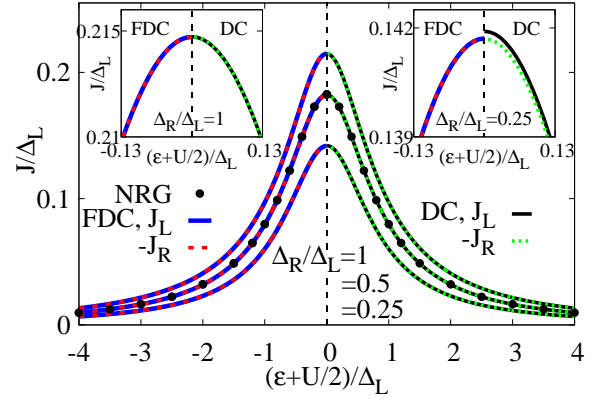}\caption{(Color online) Supercurrent through the left/right junctions as a
function of $\varepsilon$ for $U=\Delta_{L}$, $\Phi=\pi/2$, $\Gamma_{L}=2\Gamma_{R}=\Delta_{L}/2$,
and $\Delta_{R}/\Delta_{L}=1,1/2,1/4$. The current was calculated
using the FDC approximation for $\varepsilon+U/2<0$ and DC approximation
for $\varepsilon+U/2\geq0$. The insets show details close to the
half-filling. \label{fig:Current}}
\end{figure}

In Appendix \ref{App:Charge-conservation}, we have proven that the
FDC approximation is charge conserving in the general case and that DC approximation
is charge conserving for the experimentally generic case of equivalent gaps.  
The charge conservation for the DC unfortunately does not extend to the general case, giving it the same status as the conventional implementation of fRG.\cite{Karrasch-PhD10}  
In Fig.~\ref{fig:Current}, we compare
the numerical results obtained by these two methods with the NRG.
The supercurrents at the left/right junctions are plotted as functions
of $\varepsilon$ for $U=\Delta_{L}$, $\Phi=\pi/2$, $\Gamma_{L}=2\Gamma_{R}=\Delta_{L}/2$
and three values of $\Delta_{R}/\Delta_{L}$. We used the FDC
approximation to calculate the supercurrent for $\varepsilon+U/2<0$
and the DC approximation for $\varepsilon+U/2\geq0$. It can be seen
that both approaches are in excellent agreement with the NRG. However,
only the FDC approximation fully conserves the current in the general
case $\Delta_{L}\neq\Delta_{R}$. This is shown in the insets of Fig.~\ref{fig:Current}
where the details of the current are plotted close to half-filling.
The DC approximation conserves current for $\Delta_{L}=\Delta_{R}$
but not for $\Delta_{L}\neq\Delta_{R}$ as illustrated in the right
inset. The difference between $J_{L}$ and $-J_{R}$ is nevertheless
very small, below 1\% for the used parameters (see the vertical $J$-axis
scale in the inset), which justifies the widespread usage of the DC
approximation in this work as a faster and numerically more stable
(especially around the phase boundary) alternative to the FDC approximation,
which still gives trustworthy results even for $\Delta_{L}\neq\Delta_{R}$.

\begin{figure}
\includegraphics[width=1\columnwidth]{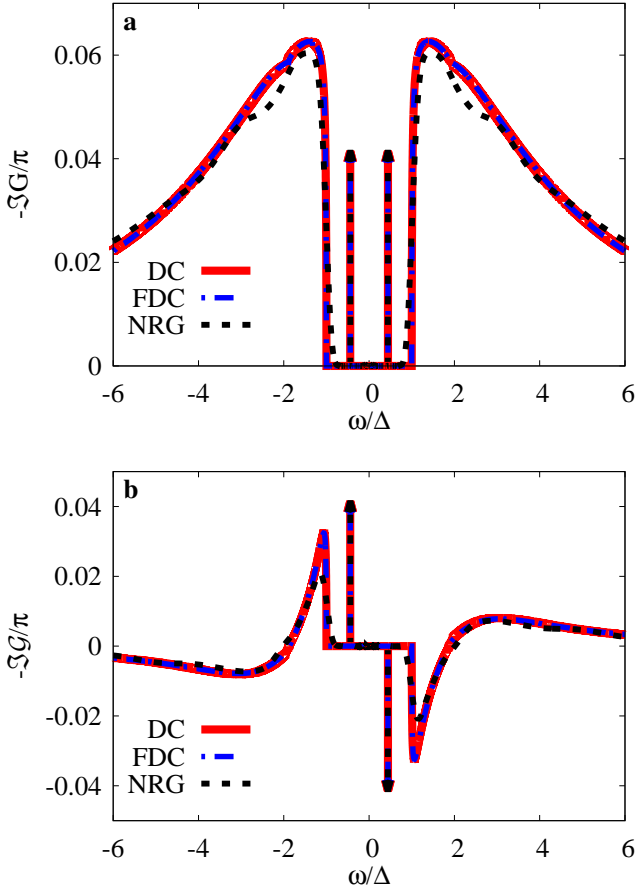} \caption{(Color online) Normal $-\Im G/\pi$ \textbf{(a)}
and anomalous $-\Im\mathcal{G}/\pi$ \textbf{(b)}
spectral densities for $U=4\Delta$, $\Gamma_R=\Delta$, $\Gamma_L=2\Gamma_{R}$ (asymmetric coupling), $\Phi=\pi/2$,
and $\varepsilon=-U/2$ (half-filling) calculated using the DC approximation
(red solid line), FDC approximation (blue dashed-dotted line) and NRG (black dashed line). Discrete Andreev bound
states within the gap are represented by arrows whose height is determined
by their weight. \label{fig:Spec}}
\end{figure}

Finally, we mention yet another aspect of the DC method, which is
its ability to calculate also spectral functions. This is possible
by making use of analytic continuation of the Matsubara formalism
to the real frequencies as shown in Refs.~\onlinecite{Zonda15,Janis15,Janis14}.
Here we just plot  the typical normal and anomalous
spectral densities for the asymmetric coupling to the leads 
calculated using the DC approximation (solid red line), FDC approximation (dashed-dotted blue line)
and NRG (dashed black line) in Fig.~\ref{fig:Spec}. Discretization parameter $\Lambda=4$ and
the logarithmic-exponential broadening of the data with the broadening
parameter $b=0.15$ together with the $z$-trick (see the manual to
Ref.~\onlinecite{NRGLjubljana}) was used for the NRG plot. Considering
the simplicity of the (F)DC approximations and the broadening of NRG curves,
which is fully responsible for the discrepancies around the band-edge,
the spectral functions are in a very good agreement, notice especially
the perfect agreement both in the position and weight of the ABS.      

\section{Comparison with experiments\label{sec:Experiments}}

\subsection{Grenoble experiment {[}Phys. Rev. X 2, 011009 (2012){]}}

\begin{figure}
\includegraphics[width=1\columnwidth]{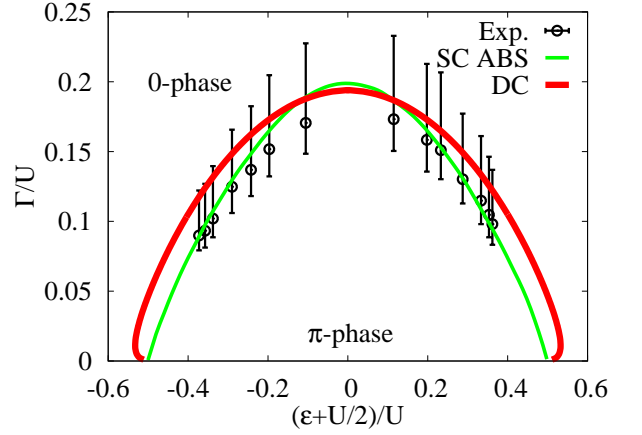}\caption{(Color online) Phase boundary between $0$- and $\pi$-phase as a
function of $\varepsilon$ and $\Gamma/U$. Both the experimental
data (with the estimated Coulomb repulsion $U=0.8\,$ meV and SC gap
$\Delta=0.08$ meV) and the theoretical curve calculated using the
self-consistent description of Andreev bound states (SC ABS)\cite{Meng09}
were taken graphically from Fig.~6 of Ref.~\onlinecite{Maurand12}. The
DC phase boundary was calculated for $U/\Delta=10$ and $\Phi=0$.\label{fig:Experiment1}}
\end{figure}

Realization of a fully tunable superconducting carbon-nanotube quantum
dot SQUID by Maurand \emph{et al.}\cite{Maurand12} allowed to determine
the phase diagram for the $0-\pi$ phase transition in the $\Gamma-\varepsilon$
plane experimentally (Fig.~\ref{fig:Experiment1}). In the experiment
the Coulomb interaction was measured from the finite-bias spectroscopy
of the Coulomb-blockade diamonds and estimated to be $U=0.80\pm0.05$
meV. The superconducting gap $\Delta\approx0.08$ meV was determined
from the peaks in the nonequilibrium (finite-bias) differential conductance.
The analysis of the maximum of normal-state conductance showed that
the tunneling amplitudes to the leads were balanced --- therefore,
the symmetric setup ($\Gamma/2=\Gamma_{L}=\Gamma_{R}$) was assumed.
Hybridization $\Gamma$ for different tunings of the setup was estimated
from the half-width at half-maximum of the Kondo resonance in the
finite-bias conductance. The authors argue that the Kondo screening
plays a key role for the $0-\pi$ phase transition in their device.
This statement is supported by a quantitative comparison
of the position of phase boundary with their theoretical predictions
based on the SC ABS approximation,\cite{Meng09} which is a perturbative method based on the superconducting
atomic limit\cite{Bauer07,Karrasch08} with expansion for finite gap
$\Delta$. The key role of the Kondo screening is emphasized also from
the identified operating regime of the experiment marked in Fig.~7
of Ref.~\onlinecite{Maurand12}. Considering this and the limitations of
the DC approach discussed above it is surprising that the actual DC
phase boundary calculated for the same parameters as SC ABS ($U/\Delta=10$)
is very close to the experimental one as shown in Fig.~\ref{fig:Experiment1}.
Although the SC ABS phase boundary is closer to experimental points
than the DC boundary, that is still within the experimental error
bars. Therefore, we conclude that the DC method performs well even
beyond its expected validity range and can be applied to a very broad
range of real superconducting quantum dots. 

\subsection{Orsay experiment {[}Phys. Rev. B 91, 241401(R) (2015){]}}

The most recent experimental study\cite{Delagrange15} of the superconducting
carbon-nanotube quantum dot confirmed theoretical predictions that
the $0-\pi$ phase transition can be controlled not only by the gate
voltage but also by the superconducting phase difference $\Phi$ tuned
by the magnetic flux piercing the SQUID loop with the carbon-nanotube
Josephson junction. The superconducting gap of the leads $\Delta=0.17$
meV and the Coulomb interaction $U=3.2$ meV, both with uncertainty
$\sim 10\%$, were experimentally determined by standard
methods (see the previous subsection). Total hybridization $\Gamma=\Gamma_{R}+\Gamma_{L}=0.44$
meV and the asymmetry $\Gamma_{R}/\Gamma_{L}=4$ of the couplings
were obtained by comparing the measured normal-state finite-temperature
linear conductance with the one obtained from CT-INT quantum Monte
Carlo\cite{Luitz10} calculations for the Anderson impurity model
analogously to Ref.~\onlinecite{Luitz12}.  The experimental results were
compared to QMC calculations and an excellent agreement was observed
both for the current-phase relation as well as for the shape and width
of the $0-\pi$ boundary in the $\Phi-\varepsilon$ plane. However,
a shift of the energy level $\delta\varepsilon=0.28$ meV of unknown
origin was needed to overlap the experimental data with the theoretical
curve.

\begin{figure}
\includegraphics[width=1\columnwidth]{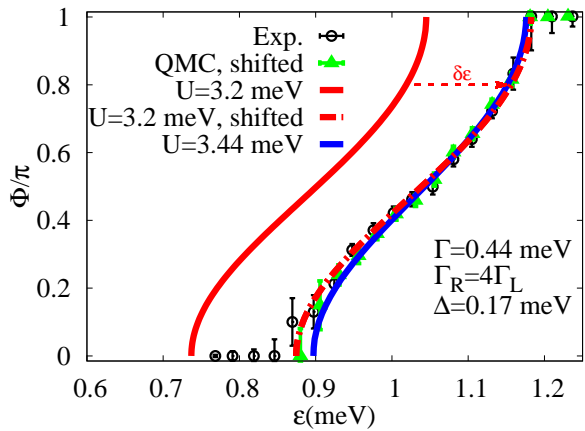}\caption{(Color online) Comparison of phase boundary in the $\Phi-\varepsilon$
plane obtained experimentally with different theoretical predictions.
Both the experimental data and QMC points were taken graphically from
Fig.~4b in Ref.~\onlinecite{Delagrange15}. Experimentally determined
parameters (for details see the main text) $\Delta=0.17$ meV, $U=3.2$
meV, $\Gamma=\Gamma_{R}+\Gamma_{L}=0.44$ meV, and $\Gamma_{R}/\Gamma_{L}=4$
are subject to roughly $10\%$ uncertainty. According to Ref.~\onlinecite{Delagrange15}
the QMC curve was horizontally shifted to match the experiment. We
plot the DC approximation for $U=3.2\,$meV (solid red curve), the
same result shifted by $\delta\varepsilon$ to overlap the experiment
(dashed red curve), and result for $U=3.44$ meV within the experimental
uncertainty without any further modification (blue full line). \label{fig:Experiment2}}
\end{figure}

In Fig.~\ref{fig:Experiment2} we have calculated the $0$-phase
boundary with the DC approximation (solid red line) and compared it
with the experimental and QMC data taken graphically from Fig.~4b
in Ref.~\onlinecite{Delagrange15}. One can see that after introducing
a small shift $\delta\varepsilon=0.14$ meV, exactly as it was done
for the QMC results, the simple second order perturbation theory reproduces
(the shape and width of) the phase boundary almost perfectly (dashed-dotted
red curve in Fig.~\ref{fig:Experiment2}). We have taken the advantages
of the DC approximation (its simplicity and speed) to check how the
boundary depends on the variance of used parameters. We have found
out, that although the shape and width of the boundary are quite robust
within the 10\% uncertainty, the $\varepsilon$-position of the boundary
is very sensitive to the value of $U$. No shift of the phase boundary
is needed if the value $U=3.44$ meV within the $U$-uncertainty range
is used instead as it is shown in Fig.~\ref{fig:Experiment2} (blue
solid curve). This leads us to the conclusion that the deviations
between theory and experiment observed in Ref.~\onlinecite{Delagrange15}
are within the experimental uncertainty. Moreover, this is yet another
demonstration of the usefulness of the DC approach for real systems.

\section{Conclusions\label{sec:Conclusions}}

To summarize, we presented a detailed study of the self-consistent
second-order perturbation expansion in the interaction strength of
the superconducting single-impurity Anderson model. Based on a thorough
analysis of its properties, we showed that it can reliably substitute
time and resources consuming  numerical methods such us the NRG or QMC for the study
of the $0-\pi$ phase transitions and properties of the $0$-phase
in superconducting quantum dots for a broad range of parameters. It
can be the method of first choice for realistic setups with asymmetric
tunnel couplings and even for unconventional setups with different
SC leads. We disclosed its big potential by successful fits
of two existing experimental data sets for the $0-\pi$ phase boundary,
including the suggestion for a plausible explanation of the existing
discrepancy between the newest experiment and corresponding QMC results.

The approach can be straightforwardly applied to any single-particle
quantity in the $0$ phase such as supercurrent, local occupation
and proximity-induced superconducting gap, or energies and weights
of the Andreev bound states including the position of the $0-\pi$
quantum phase boundary at zero temperature. Due to its perturbation-theory
roots it, however, conceptually fails in the description of the $\pi$ phase
and, consequently, also in the description of the finite-temperature
properties close to the phase boundary. A possible remedy of the perturbation
approach to reach also the $\pi$ phase with the doublet ground state
remains an open challenge, which in view of the successes of the method
in the 0-phase is worth taking up in future studies. 

\begin{acknowledgments}
This work was supported by Grant No.~15-08740Y of the Czech Science
Foundation (M.\v{Z}.) and by the National Science Centre (Poland) under
the contract DEC-2014/13/B/ST3/04451 (T.N.). V.J.\ thanks the Fulbright Commission for financing his stay at Louisiana State University where part of the research was performed. Access to computing and storage facilities owned by parties and projects contributing to the
National Grid Infrastructure MetaCentrum, provided under the program
``Projects of Large Infrastructure for Research, Development, and
Innovations'' (LM2010005), is greatly appreciated.
\end{acknowledgments}

\appendix

\section{Hartree-Fock phase boundary\label{App:Hartree-Fock}}

Initial version of HF equations \eqref{eq:HFinit} can be recast into
the simpler form \eqref{eq:HFreform} by introducing auxiliary quantities
$E_{d}\equiv\Sigma^{\mathrm{HF}}+\varepsilon$ and $\delta\equiv\sum_{\alpha=L,R}\Gamma_{\alpha}e^{i\Phi_{\alpha}}-\mathcal{S}^{\mathrm{HF}}$
and further using the general identity 
\begin{equation}
-\frac{1}{\beta}\sum_{\omega_{n}}e^{i\omega_{n}0^{+}}\frac{i\omega_{n}[1+s(i\omega_{n})]}{D(i\omega_{n})}=\frac{1}{2}\label{eq:sum-rule}
\end{equation}
valid for \emph{any} spin-symmetric GF \eqref{eq:GF} due to a sum
rule reflecting the fundamental anticommutation relation
\begin{widetext}
\begin{equation}
\begin{split}\frac{1}{\beta}\sum_{\omega_{n}}e^{i\omega_{n}0^{+}}[G(i\omega_{n})+\bar{G}(i\omega_{n})] & =-\frac{1}{\beta}\sum_{\omega_{n}}e^{i\omega_{n}0^{+}}\frac{2i\omega_{n}[1+s(i\omega_{n})]+\Sigma^{*}(i\omega_{n})-\Sigma(i\omega_{n})}{D(i\omega_{n})}\\
 & =G(\tau-\tau'\to0^{-})+\bar{G}(\tau-\tau'\to0^{-})\equiv<d^{\dagger}d>+<dd^{\dagger}>=1.
\end{split}
\end{equation}

\end{widetext}

Large-frequency behavior of the self-energy $\Sigma(i\omega_{n})$
is limited by a constant (in case of the HF approximation; otherwise
it generically decays as $1/i\omega_{n}$), which allows us to drop
the phase-convergence factor in the above sum and using the symmetry
relations $\Sigma^{*}(i\omega_{n})=\Sigma(-i\omega_{n})$ (implying
real $\Sigma^{\mathrm{HF}}$) and $D(-i\omega_{n})=D(i\omega_{n})$
\eqref{eq:Det} we get $\sum_{\omega_{n}}e^{-i\omega_{n}0^{+}}\frac{\Sigma^{*}(i\omega_{n})-\Sigma(i\omega_{n})}{D(i\omega_{n})}=\sum_{\omega_{n}}\frac{\Sigma(-i\omega_{n})-\Sigma(i\omega_{n})}{D(i\omega_{n})}=0$
thus proving the required identity \eqref{eq:sum-rule}.

Determinant $D^{\mathrm{HF}}(i\omega_{n})$ explicitly reads
\begin{widetext}
\begin{gather}
\begin{split}D^{\mathrm{HF}}(i\omega_{n}) & =\omega_{n}^{2}\left[1+\sum_{\alpha}\frac{\Gamma_{\alpha}}{\sqrt{\Delta_{\alpha}^{2}+\omega_{n}^{2}}}\right]^{2}+\left[\varepsilon+\Sigma^{\mathrm{HF}}\right]^{2}+\left|\sum_{\alpha}\frac{\Gamma_{\alpha}\Delta_{\alpha}}{\sqrt{\Delta_{\alpha}^{2}+\omega_{n}^{2}}}e^{i\Phi_{\alpha}}-\mathcal{S}^{\mathrm{HF}}\right|^{2}\\
 & =\omega_{n}^{2}\left[1+\sum_{\alpha}\frac{\Gamma_{\alpha}}{\sqrt{\Delta_{\alpha}^{2}+\omega_{n}^{2}}}\right]^{2}+E_{d}^{2}+\left|\sum_{\alpha}\Gamma_{\alpha}e^{i\Phi_{\alpha}}\left(\frac{\Delta_{\alpha}}{\sqrt{\Delta_{\alpha}^{2}+\omega_{n}^{2}}}-1\right)+\delta\right|^{2}\\
 & \approx E_{d}^{2}+|\delta|^{2}+\left[\left(1+\sum_{\alpha}\frac{\Gamma_{\alpha}}{\Delta_{\alpha}}\right)^{2}-\sum_{\alpha}\frac{\Gamma_{\alpha}}{\Delta_{\alpha}^{2}}\Re\left(\delta e^{-i\Phi_{\alpha}}\right)\right]\omega_{n}^{2}.
\end{split}
\label{eq:DHF}
\end{gather}
\end{widetext}
Because close to the QPT both $E_{d}$ and $\delta$ are close to
zero, the second term in the brackets multiplying $\omega_{n}^{2}$
can be safely neglected in the calculation of the phase boundary (as
is done in the main text) since this term is effectively of higher
order in the $\omega_{n}$-expansion. 

Finally, we discuss the band contribution term $\mathcal{B}$ \eqref{eq:Band}.
Its name derives from the fact that when the integral over the Matsubara
frequencies \eqref{eq:Band} is Wick-rotated to the real frequencies,
it only contains the continuous (band) part of the spectrum, i.e.,
it does not encompass any ABS contributions. The general formula can
be recast into a more compact form for the generic case with equal
SC gaps $\Delta_{L}=\Delta_{R}=\Delta$. Using the substitution $\omega=\Delta\sinh t$
and mutually canceling the common $2\sinh^{2}\frac{t}{2}$ terms in
the numerator and denominator of the integrand, we arrive at the expression
\begin{widetext}
\begin{equation}
\mathcal{B}=\frac{\sum_{\alpha}\Gamma_{\alpha}e^{i\Phi_{\alpha}}}{\Delta}\int_{0}^{2\pi}\frac{dt}{2\pi}\frac{\cosh^{2}t}{\left(\cosh t+\frac{\sum_{\alpha}\Gamma_{\alpha}}{\Delta}\right)^{2}\cosh^{2}\frac{t}{2}+\left|\frac{\sum_{\alpha}\Gamma_{\alpha}e^{i\Phi_{\alpha}}}{\Delta}\right|^{2}\sinh^{2}\frac{t}{2}}
\end{equation}
\end{widetext}
which generalizes the symmetric case $\Gamma_{L}=\Gamma_{R}=\Gamma/2$
(and $\Phi_{L}=-\Phi_{R}=\Phi/2$) studied previously in Ref.~\onlinecite{Zonda15}.

\section{Charge conservation for the dynamical corrections\label{App:Charge-conservation}}

\subsection{FDC}
As a special case of Eq.~\eqref{eq:charge-cons} we now consider
the charge conservation in the second-order approximation for which
the ``current defect'' reads
\begin{equation}
\delta J^{(2)}=-\frac{4}{\beta}\Im\sum_{\omega_{n}}\mathcal{S}^{(2)}(i\omega_{n})\mathcal{G}^{*}(i\omega_{n})\label{eq:dJ2}
\end{equation}
with the anomalous self-energy \eqref{eq:2ndsean}
\begin{align}
\mathcal{S}^{(2)}(i\omega_{n}) & =-\frac{U^{2}}{\beta}\sum_{\nu_{m}}\mathcal{G}(i\omega_{n}+i\nu_{m})\chi(i\nu_{m}),
\end{align}
and the bubble contribution \eqref{eq:bubble}
\begin{gather}
\chi(i\nu_{m})=\frac{1}{\beta}\sum_{\omega_{k}}\left[G(i\nu_{m}+i\omega_{k})G(i\omega_{k})+\mathcal{G}(i\nu_{m}+i\omega_{k})\mathcal{G}^{*}(i\omega_{k})\right].
\end{gather}
Separating the quantity $\delta J^{(2)}=\delta J_{n}^{(2)}+\delta J_{a}^{(2)}$
into two parts corresponding to the normal and anomalous Green functions
constituents of the bubble, respectively, we get
\begin{widetext}
\begin{equation}
\begin{split}\delta J_{n}^{(2)} & =\frac{4U^{2}}{\beta^{3}}\Im\sum_{\omega_{n},\omega_{k},\nu_{m}}G(i\nu_{m}+i\omega_{k})G(i\omega_{k})\mathcal{G}(i\omega_{n}+i\nu_{m})\mathcal{G}^{*}(i\omega_{n}),\\
\delta J_{a}^{(2)} & =\frac{4U^{2}}{\beta^{3}}\Im\sum_{\omega_{n},\omega_{k},\nu_{m}}\mathcal{G}(i\nu_{m}+i\omega_{k})\mathcal{G}^{*}(i\omega_{k})\mathcal{G}(i\omega_{n}+i\nu_{m})\mathcal{G}^{*}(i\omega_{n}).
\end{split}
\end{equation}
Using the symmetry relation $G^{*}(i\omega)=G(-i\omega)$ we can manipulate
the first formula as
\begin{equation}
\begin{split}\delta J_{n}^{(2)} & =\frac{2U^{2}}{\beta^{3}}\sum_{\omega_{n},\omega_{k},\nu_{m}}\left[G(i\nu_{m}+i\omega_{k})G(i\omega_{k})\mathcal{G}(i\omega_{n}+i\nu_{m})\mathcal{G}^{*}(i\omega_{n})-G^{*}(i\nu_{m}+i\omega_{k})G^{*}(i\omega_{k})\mathcal{G}^{*}(i\omega_{n}+i\nu_{m})\mathcal{G}(i\omega_{n})\right]\\
 & =\frac{2U^{2}}{\beta^{3}}\sum_{\omega_{n},\omega_{k},\nu_{m}}\left[G(i\nu_{m}+i\omega_{k})G(i\omega_{k})\mathcal{G}(i\omega_{n}+i\nu_{m})\mathcal{G}^{*}(i\omega_{n})-G(-i\nu_{m}-i\omega_{k})G(-i\omega_{k})\mathcal{G}^{*}(i\omega_{n}+i\nu_{m})\mathcal{G}(i\omega_{n})\right]\\
 & =\frac{2U^{2}}{\beta^{3}}\sum_{\omega_{n},\omega_{k},\nu_{m}}\left[G(i\nu_{m}+i\omega_{k})G(i\omega_{k})\mathcal{G}(i\omega_{n}+i\nu_{m})\mathcal{G}^{*}(i\omega_{n})-G(i\nu_{m}+i\omega_{k})G(i\omega_{k})\mathcal{G}^{*}(i\omega_{n}-i\nu_{m})\mathcal{G}(i\omega_{n})\right]\\
 & =0,
\end{split}
\end{equation}
\end{widetext}
where we have used substitutions $\omega_{k}\rightarrow-\omega_{k}$
and $\nu_{m}\rightarrow-\nu_{m}$ in the second term of the sums between
the second and the third lines and then the shift of the summation
variable $\omega_{n}-\nu_{m}\rightarrow\omega_{n}$ in the last step.
Analogously, the anomalous contribution can be simplified with help
of the substitution $\nu_{m}\rightarrow-\nu_{m}$ and shift of variables
$\omega_{n,k}\rightarrow\omega_{n,k}+\nu_{m}$ as follows:
\begin{widetext}
\begin{equation}
\begin{split}\delta J_{a}^{(2)} & =\frac{2U^{2}}{\beta^{3}}\sum_{\omega_{n},\omega_{k},\nu_{m}}\left[\mathcal{G}(i\nu_{m}+i\omega_{k})\mathcal{G}^{*}(i\omega_{k})\mathcal{G}(i\omega_{n}+i\nu_{m})\mathcal{G}^{*}(i\omega_{n})-\mathcal{G}^{*}(i\nu_{m}+i\omega_{k})\mathcal{G}(i\omega_{k})\mathcal{G}^{*}(i\omega_{n}+i\nu_{m})\mathcal{G}(i\omega_{n})\right]\\
 & =\frac{2U^{2}}{\beta^{3}}\sum_{\omega_{n},\omega_{k},\nu_{m}}\left[\mathcal{G}(i\nu_{m}+i\omega_{k})\mathcal{G}^{*}(i\omega_{k})\mathcal{G}(i\omega_{n}+i\nu_{m})\mathcal{G}^{*}(i\omega_{n})-\mathcal{G}^{*}(-i\nu_{m}+i\omega_{k})\mathcal{G}(i\omega_{k})\mathcal{G}^{*}(i\omega_{n}-i\nu_{m})\mathcal{G}(i\omega_{n})\right]\\
 & =0,
\end{split}
\end{equation}
\end{widetext}
which finalizes the required proof of the conserving nature $\delta J^{(2)}=0$
of the FDC approximation.

\subsection{DC}
As we have shown numerically in Sec.~\ref{sub:Applicability} the DC approximation is charge conserving for identical gaps 
$\Delta_L=\Delta_L=\Delta$. This can be proven analytically by showing that both 
$\mathcal{G}^\mathrm{DC}(i\omega_{n})$ and $\mathcal{S}^\mathrm{DC}(i\omega_{n})$ are real which is a sufficient condition 
for $\delta J=0$ in Eq.~\eqref{eq:dJ2}. By making use of the gauge invariance it is possible to introduce a {\em global} phase shift 
\begin{equation}
\phi_{\mathrm{sh}}=\arctan\left(\frac{\Gamma_L - \Gamma_R}{\Gamma_L + \Gamma_R}\tan\frac{\phi}{2}\right),
\label{eq:phaseshift}
\end{equation}
such that $\phi_L=\phi_{\mathrm{sh}}-\phi/2$, $\phi_R=\phi_{\mathrm{sh}}+\phi/2$
for which $\Delta_{\phi}(i\omega_{n})$ is real for all frequencies. Consequently, the  $\mathcal{G}^\mathrm{HF}(i\omega_{n})$ and $\mathcal{S}^\mathrm{HF}$ are real too, because the equality 
\begin{equation}
2\Im \mathcal{S}^\mathrm{HF}=\mathcal{S}^\mathrm{HF}-{\mathcal{S}^\mathrm{HF}}^*=
-2\Im \mathcal{S}^\mathrm{HF}\frac{U}{\beta}\sum_{\omega_{n}}\frac{1}{D^{\mathrm{HF}}(i\omega_{n})}
\label{eq:SHFreal}
\end{equation}
can be generally fulfilled only when $\Im \mathcal{S}^\mathrm{HF}=0$. 
Assuming the contrary, i.e., the existence of the special solution 
\begin{equation}
\frac{U}{\beta}\sum_{\omega_{n}}\frac{1}{D^{\mathrm{HF}}(i\omega_{n})}=-1
\label{eq:SHFcon}
\end{equation} 
implies from Eq.~\eqref{eq:HFinit} the identity $\frac{U}{\beta}\sum_{\omega_{n}}\frac{\Delta_\phi (i\omega_{n})}{D^{\mathrm{HF}}(i\omega_{n})}=0$ which would
in turn mean that Eq.~\eqref{eq:HFse} is fulfilled for {\em any} $\mathcal{S}^\mathrm{HF}$. Correspondingly, the same
must be true also for Eq.~\eqref{eq:SHFcon} which can be easily contradicted, e.g., by taking limit $\mathcal{S}^\mathrm{HF}\to\infty$.     
Reality of $\mathcal{G}^\mathrm{HF}(i\omega_{n})$ then follows directly from the Eq.~\eqref{eq:GFHF}. 

The second-order contribution to the anomalous DC self-energy is also real because it reads as
\begin{equation}
\mathcal{S}^{(2)}(i\omega_{n}) =-\frac{U^{2}}{\beta}\sum_{\nu_{m}}\mathcal{G}^\mathrm{HF}(i\omega_{n}+i\nu_{m})\chi^\mathrm{HF}(i\nu_{m}),
\end{equation}
where both $\mathcal{G}^\mathrm{HF}(i\omega_{n})$ and the bubble contribution $\chi^\mathrm{HF}(i\nu_{m})$ (see Eq.~\eqref{eq:bubble}) are real.
However, the first order contribution to the anomalous DC self energy reads as ${\mathcal{S}^{(1)}}=\frac{U}{\beta}\sum_{\omega_{n}}\mathcal{G}^\mathrm{DC}(i\omega_{n})$ where
\begin{equation}
\mathcal{G}^\mathrm{DC}(i\omega_{n})=-\frac{1}{D^{\mathrm{DC}}(i\omega_{n})}\left(\mathcal{S}^{(1)}+\mathcal{S}^{(2)}(i\omega_{n})-\Delta_\phi(i\omega_{n})\right)
\end{equation}
with all $\mathcal{S}^{(2)}(i\omega_{n})$, $D^{\mathrm{DC}}(i\omega_{n})$, $\Delta_\phi(i\omega_{n})$ being real. 
Therefore, using the same argument as for $S^\mathrm{HF}$ and $\mathcal{G}^\mathrm{HF}(i\omega_{n})$, one can show that $\mathcal{S}^{(1)}$ is real. 
Consequently, also $S^{\mathrm{DC}}=\mathcal{S}^{(1)}+\mathcal{S}^{(2)}$ and $\mathcal{G}^\mathrm{DC}(i\omega_{n})$ are real which was to be proven. 

Note that this proof does not carry over to the general case $\Delta_{L}\neq\Delta_{R}$ due to the lack of existence of a global, i.e., frequency-independent phase shift to make $\Delta_{\phi}(i\omega_{n})$ real for all $\omega_{n}$'s. The DC approximation is thus not conserving for non-identical leads as revealed in our numerical results,  although the observed current conservation breaking is very weak (see Fig.~ \ref{fig:Current}).       

%\bibliographystyle{apsrev4-1}
%\bibliography{Josephson}

%

\end{document}